\documentclass{jpp}
\usepackage{graphicx,natbib,jab,upmath,amssymb,amsbsy}

\title[Solar wind turbulence]{Recent progress in astrophysical plasma turbulence from solar wind observations}
\author[C. H. K. Chen]{C.\ns H.\ns K.\ns C\ls H\ls E\ls N$^1$\thanks{Email address for correspondence: christopher.chen@imperial.ac.uk}}
\affiliation{$^1$Department of Physics, Imperial College London, London SW7 2AZ, UK}
\begin{document}
\maketitle
\begin{abstract}
This paper summarises some of the recent progress that has been made in understanding astrophysical plasma turbulence in the solar wind, from \emph{in situ} spacecraft observations. At large scales, where the turbulence is predominantly Alfv\'enic, measurements of critical balance, residual energy, and 3D structure are discussed, along with comparison to recent models of strong Alfv\'enic turbulence. At these scales, a few percent of the energy is also in compressive fluctuations, and their nature, anisotropy, and relation to the Alfv\'enic component is described. In the small scale kinetic range, below the ion gyroscale, the turbulence becomes predominantly kinetic Alfv\'en in nature, and measurements of the spectra, anisotropy, and intermittency of this turbulence are discussed with respect to recent cascade models. One of the major remaining questions is how the turbulent energy is dissipated, and some recent work on this question, in addition to future space missions which will help to answer it, are briefly discussed.
\end{abstract}

\section{Introduction}

Plasma turbulence is one of the most widespread collective phenomena occurring in nature. It appears to be present throughout the universe, in a diverse range of environments, including galaxy clusters, accretion disks, the interstellar medium, stars, stellar winds, and planetary magnetospheres. While being an intriguing and complex aspect of our universe of intrinsic interest, it can also have an important impact on the large-scale properties of these systems. For example, it can enable angular momentum transport in accretion disks \citep{balbus98}, galactic magnetic field amplification \citep{kulsrud08}, limit thermal conduction in galaxy clusters \citep{schekochihin08b}, determine the dispersal and mixing of elements in the interstellar medium \citep{scalo04}, and play a key role in star formation \citep{mckee07}. In addition, the dissipation of turbulent energy may explain the large temperatures observed in many astrophysical systems, such as the solar corona \citep{cranmer15}, and galaxy clusters \citep{zhuravleva14short}. Turbulent plasmas display complex, chaotic, broadband fluctuations, which are generally interpreted as a scale-invariant cascade of energy from large scales, where the energy is injected, to small scales, where it is dissipated. However, many questions remain about both the cascade and dissipation processes.

The solar wind presents one of the best places to understand the basic physics of plasma turbulence. The main reason for this is that \emph{in situ} spacecraft measurements are possible, meaning that a wealth of information is available, unlike for plasmas outside our solar system. It is also fast flowing, travelling much quicker than the local Alfv\'en speed and turbulent fluctuation amplitudes, meaning that time series measured by spacecraft can generally be interpreted as spatial cuts though the plasma (the Taylor hypothesis), enabling them to be readily compared to theoretical predictions. The majority of solar wind turbulence measurements, and most of those described in this paper, have been made in the near-Earth solar wind. Typical plasma conditions here\footnote{As measured by the \emph{Wind} spacecraft using the data set described in \citet{chen16a}.} are: magnetic field strength $B\sim 6\,\textrm{nT}$, number density $n_\mathrm{i}\sim n_\mathrm{e}\sim 11\,\textrm{cm}^{-3}$, bulk speed $v_\mathrm{i}\sim v_\mathrm{e}\sim 390\,\textrm{km\,s}^{-1}$, and ion and electron temperatures $T_\mathrm{i}\sim 6\,\mathrm{eV}$ and $T_\mathrm{e}\sim 12\,\textrm{eV}$, which give ion and electron plasma betas of order unity, $\beta_\mathrm{i}\sim\beta_\mathrm{e}\sim 1$. The dominant ion species (to which these parameters refer) is Hydrogen, with Helium and other minor ions making up a few per cent of the solar wind by number density.  It is important to note, however, that there is large a variation in all of these values, making the solar wind an ideal place for parameter studies of plasma processes, applicable to a wide variety of astrophysical environments.

This paper summarises some of the recent progress that has been made in understanding solar wind turbulence. It is not intended to be a complete review; for general review papers on solar wind observations see, e.g., \citet{alexandrova13a}, \citet{bruno13}, and the collection of \citet{kiyani15}. It also primarily summarises work in which I have been involved, with other results and theoretical background described for context. Therefore, it is not an unbiased review, but presents a viewpoint, based on the latest observations and theoretical developments.

\section{Alfv\'enic turbulence}
\label{sec:alfvenicturbulence}

The solar wind has long been known to contain large scale Alfv\'enically polarised fluctuations, consistent with predominant propagation away from the Sun \citep[e.g.,][]{belcher71}. These are thought to be Alfv\'en waves, generated at or near the Sun, which propagate into interplanetary space and drive a turbulent cascade. The magnetic fluctuations at these large scales are often seen to have a $1/f$ frequency spectrum, which has been interpreted as arising from uncorrelated processes at the Sun \citep{matthaeus86} or a reflection driven cascade \citep{velli89,verdini12a,perez13}. Here, we are concerned with the smaller scale turbulence driven by these large scale fluctuations. In this range, the Alfv\'enically polarised fluctuations remain dominant, and in this Section our recent progress on this turbulence is discussed.

\subsection{Phenomenological models of Alfv\'enic turbulence}
\label{sec:models}

Alfv\'enic turbulence is thought to be captured by the incompressible MHD equations, which in fluctuating \citet{elsasser50} form (omitting forcing and dissipation terms) are
\begin{equation}
\label{eq:mhd}
\frac{\partial\delta\mathbf{z}^\pm}{\partial t}\mp v_\mathrm{A}\nabla_\|\delta\mathbf{z}^\pm+\delta\mathbf{z}^\mp\cdot\mathbf{\nabla}\delta\mathbf{z}^\pm=-\mathbf{\nabla}P,
\end{equation}
\begin{equation}
\nabla\cdot\delta\mathbf{z}^\pm=0,
\end{equation}
where $\delta\mathbf{z}^\pm=\delta\mathbf{v}\pm\delta\mathbf{b}$ are the fluctuating Elsasser variables, $\mathbf{v}$ is the bulk velocity, $\mathbf{b}=\mathbf{B}/\sqrt{\mu_0\rho}$ is the magnetic field in Alfv\'en units, $\rho$ is the mass density, $v_\mathrm{A}=B_0/\sqrt{\mu_0\rho}$ is the Alfv\'en speed, $P$ contains the total (thermal plus magnetic) pressure, and $\nabla_\|$ is the gradient in the direction of the mean magnetic field $\mathbf{B}_0$. An early model of Alfv\'enic turbulence, based on these equations, was developed by \citet{iroshnikov63} and \citet{kraichnan65}. From the form of the nonlinear terms ($\delta\mathbf{z}^\mp\cdot\mathbf{\nabla}\delta\mathbf{z}^\pm$) it was realised that nonlinear interactions occur via oppositely propagating Alfv\'en wave packets ($\delta\mathbf{z}^+$ and $\delta\mathbf{z}^-$). The turbulence was assumed to be isotropic so that the presence of a strong mean magnetic field would lead to weak interactions (the linear terms dominating in Equation \ref{eq:mhd}), requiring many such interactions to transfer the energy to a smaller scale. When a constant energy flux through scales is assumed, scaling arguments based on the cascade time lead to an inertial range total energy spectrum
\begin{equation}
E(k)\propto k^{-3/2},
\end{equation}
where $k$ is the wavenumber of the fluctuations.

Weak Alfv\'enic turbulence, however, was later shown to develop strong anisotropy, with a spectrum perpendicular to mean magnetic field of $E(k_\perp)\propto k_\perp^{-2}$ and no transfer in the parallel direction \citep{montgomery81,shebalin83,goldreich97,galtier00}. This leads to a violation of both the isotropic and weak assumptions as the cascade proceeds to smaller scales, since the degree of non-linearity increases with wavenumber. \citet{goldreich95} proposed that a state of critical balance is reached in which the linear and nonlinear terms become (and remain) comparable, and the transfer to smaller scales occurs within a single interaction. From Equation \ref{eq:mhd}, the linear Alfv\'en time is $\tau_\mathrm{A}\sim l_\|/v_\mathrm{A}$ (where $l_\|$ is the scale parallel to the mean magnetic field) and the nonlinear time was estimated as $\tau_\mathrm{nl}\sim l_\perp/\delta z$ (where $l_\perp$ is the perpendicular scale), which, when equated, lead to a scaling $k_\|\propto k_\perp^{2/3}$\footnote{This relation, and that fact that it leads to increasing anisotropy towards smaller scales, was also noted in the previous work of \citet{higdon84}.}. This results in an anisotropy $k_\perp\gg k_\|$ at small scales and energy spectra 
\begin{equation}
E(k_\perp) \propto k_\perp^{-5/3},\ \ \ E(k_\|) \propto k_\|^{-2}. 
\end{equation}
Since the nonlinear transfer occurs within one interaction, turbulence in critical balance is considered to be strong at all scales, even when $\delta B/B_0\ll1$. 

\citet{boldyrev06} later noted that the nonlinear time should contain an extra factor of the alignment angle $\theta$ between the fluctuations $\tau_\mathrm{nl}\sim l_\perp/(\delta z\sin\theta)$. This is because both the fluctuations and their gradients are mostly in the plane perpendicular to the mean field, and within this plane the gradients are perpendicular to the fluctuations due to solenoidality, leading to a greater reduction of the nonlinear terms if the fluctuations are more aligned. It was proposed that $\delta\mathbf{v}$ and $\delta\mathbf{b}$ align to the maximum amount permitted at each scale, leading to $\theta\propto k_\perp^{-1/4}$, and three-dimensionally anisotropic eddies at small scales $l_\|\gg\xi\gg\lambda$, where $\xi$ is the characteristic length in the fluctuation direction and $\lambda$ in the direction perpendicular to $l_\|$ and $\xi$. The resulting spectra are 
\begin{equation}
E(k_\lambda) \propto k_\lambda^{-3/2},\ \ \ E(k_\xi) \propto k_\xi^{-5/3},\ \ \ E(k_\|) \propto k_\|^{-2}.
\end{equation}
In more recent years, further additions to these models have been proposed to allow for the imbalance of the oppositely directed Alfv\'enic fluxes \citep{lithwick07,beresnyak08,chandran08,perez09,podesta10c} and intermittency \citep{chandran15,mallet16b}. While models such as these are at the phenomenological level, they provide scaling predictions that can be compared to observations to distinguish the physical processes taking place.

\subsection{Anisotropy and critical balance}
\label{sec:cb}

It is well known that Alfv\'enic turbulence in the solar wind is anisotropic \citep[see reviews by][]{horbury12,oughton15}. Correlation lengths are typically observed to be much longer in the direction parallel to the mean magnetic field than perpendicular\footnote{Although a notable exception is \citet{dasso05}, who found the opposite situation in the fast solar wind at the large scale end of the inertial range.}, consistent with the anisotropy $k_\perp\gg k_\|$ implied by the above strong turbulence models. More recently, however, new techniques have been developed to measure the anisotropic scaling, as a more direct test. In the above models, the parallel length scale is associated with the linear term, corresponding to the propagation of Alfv\'en wave packets, which are sensitive to the mean magnetic field at their location and scale, i.e., the local mean field. The need to use such a mean field to test the critical balance predictions was identified in simulations \citep{cho00,maron01} and a technique to do this in the solar wind, making use of the wavelet transform, was developed by \citet{horbury08}. At each location, and at each scale (i.e. for each wavelet coefficient in the transform), a local mean field can be defined from the average of the magnetic field weighted by the corresponding wavelet envelope. Fluctuations at a range of scales at a particular angle to the local mean field can be gathered to produce the spectrum in that direction\footnote{Such conditioning means that higher (than second) order correlations are important for the local fluctuation spectrum \citep[as noted by][]{matthaeus12a}.}. A similar technique can be applied to structure functions \citep[e.g.,][]{luo10,chen11a}, for which the local mean field can be defined as the average between the points of the structure function. In practice, the results are not sensitive to this precise definition, as long as a local (and not global) mean field is used.

\begin{figure}
\includegraphics[width=\textwidth,trim=0 0 0 0,clip]{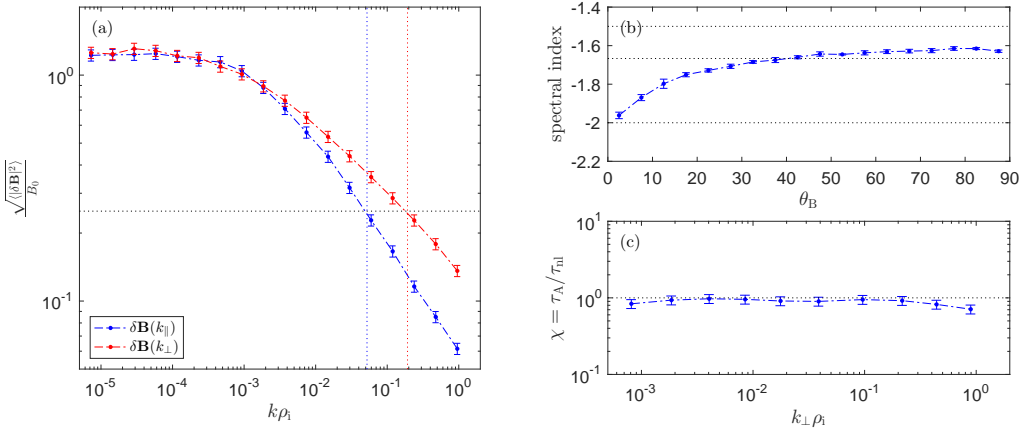}
\caption{(a) Normalised magnetic fluctuation amplitude as a function of parallel and perpendicular wavenumber. (b) Variation of spectral index with angle to the local mean field $\theta_\mathrm{B}$. (c) Ratio of linear and nonlinear time scales $\chi$ as a function of scale $k_\perp\rho_\mathrm{i}$.}
\label{fig:cb}
\end{figure}

The results of the local structure function technique \citep{chen11a}, applied to the same magnetic field data as in \citet{horbury08}, are shown in Figure \ref{fig:cb}\footnote{The error bars here represent (a) the measured statistical uncertainties on the structure functions, (b) the measured uncertainties on the fitted spectral slopes, and (c) the uncertainties from the measured structure functions propagated to $\chi$.}. It can be seen that at large scales ($k\rho_\mathrm{i}\lesssim 10^{-3}$, where $\rho_\mathrm{i}$ is the ion gyroradius) the turbulence is isotropic \citep{wicks10} with $\delta B/B_0\sim 1$, and in the inertial range ($10^{-3}\lesssim k\rho_\mathrm{i}\lesssim 1$) the parallel spectrum is steeper than the perpendicular spectrum. The inertial range spectral index as a function of angle to the local mean field, $\theta_B$, is shown in Figure \ref{fig:cb}b. The perpendicular spectrum is closer to the $k_\perp^{-5/3}$ prediction of \citet{goldreich95} than the $k_\perp^{-3/2}$ prediction of \citet{boldyrev06}, but the parallel spectrum is consistent with the $k_\|^{-2}$ critical balance prediction of both models\footnote{A recent suggestion by \citet{beresnyak15} is that the $k_\|^{-2}$ spectrum could also be interpreted as a reflection of the Lagrangian frequency spectrum.}. \citet{wicks11} also found a $k_\|^{-2}$ spectrum for the velocity and dominant Elsasser field (although instrumental noise made it difficult to measure the anisotropy of the sub-dominant Elsasser field). It is important to note that the $k_\|^{-2}$ spectrum is only observed when a scale-dependent local mean field is used; a global mean field cannot be used to test critical balance, since, even for arbitrarily small $\delta B/B_0$, the predicted anisotropy would be just large enough that the true local parallel correlation would not be measured \citep{cho00,chen11a}.

An alternative way to test the critical balance condition is to estimate the turbulence strength directly from the measured timescales. The parallel and perpendicular lengths of an ``eddy'' of amplitude $\delta B$ can be found using the technique of \citet{chen10a,chen12b}. For a particular $\delta B$, for example the horizontal dotted line in Figure \ref{fig:cb}a, $k_\|$ is estimated as the wavenumber corresponding to the parallel spectrum (blue vertical dotted line) and $k_\perp$ as the wavenumber corresponding to the perpendicular spectrum (red vertical dotted line). From this, the wavevector anisotropy can be determined, e.g., at $k_\perp\rho_\mathrm{i}=1$ the anisotropy is $k_\perp/k_\|=5.5\pm0.7$, corresponding to a wavevector angle $\theta_{kB}=(79.7\pm1.2)^{\circ}$. The ratio of linear to nonlinear times (omitting the alignment angle) is then given by $\chi=\tau_\mathrm{A}/\tau_\mathrm{nl}=(k_\perp/k_\|)(\delta B/B_0)$ and is shown in Figure \ref{fig:cb}c as a function of $k_\perp\rho_\mathrm{i}$, where it can be seen to be close to unity throughout the inertial range\footnote{Note that this is different to the result reported by \cite{wang16}, however, in that study a fixed scale-independent anisotropy was assumed in the calculation of $\chi$.}. Even though the estimate of $\tau_\mathrm{nl}$ used here is based on the magnetic fluctuations only, does not include the alignment angle, and is only defined to order unity, the critical balance condition ($\chi\sim 1$) appears to be well satisfied in the solar wind.

\subsection{Spectral indices and residual energy}
\label{sec:indices}

\begin{figure}
\includegraphics[width=0.48\textwidth,trim=0 0 0 0,clip]{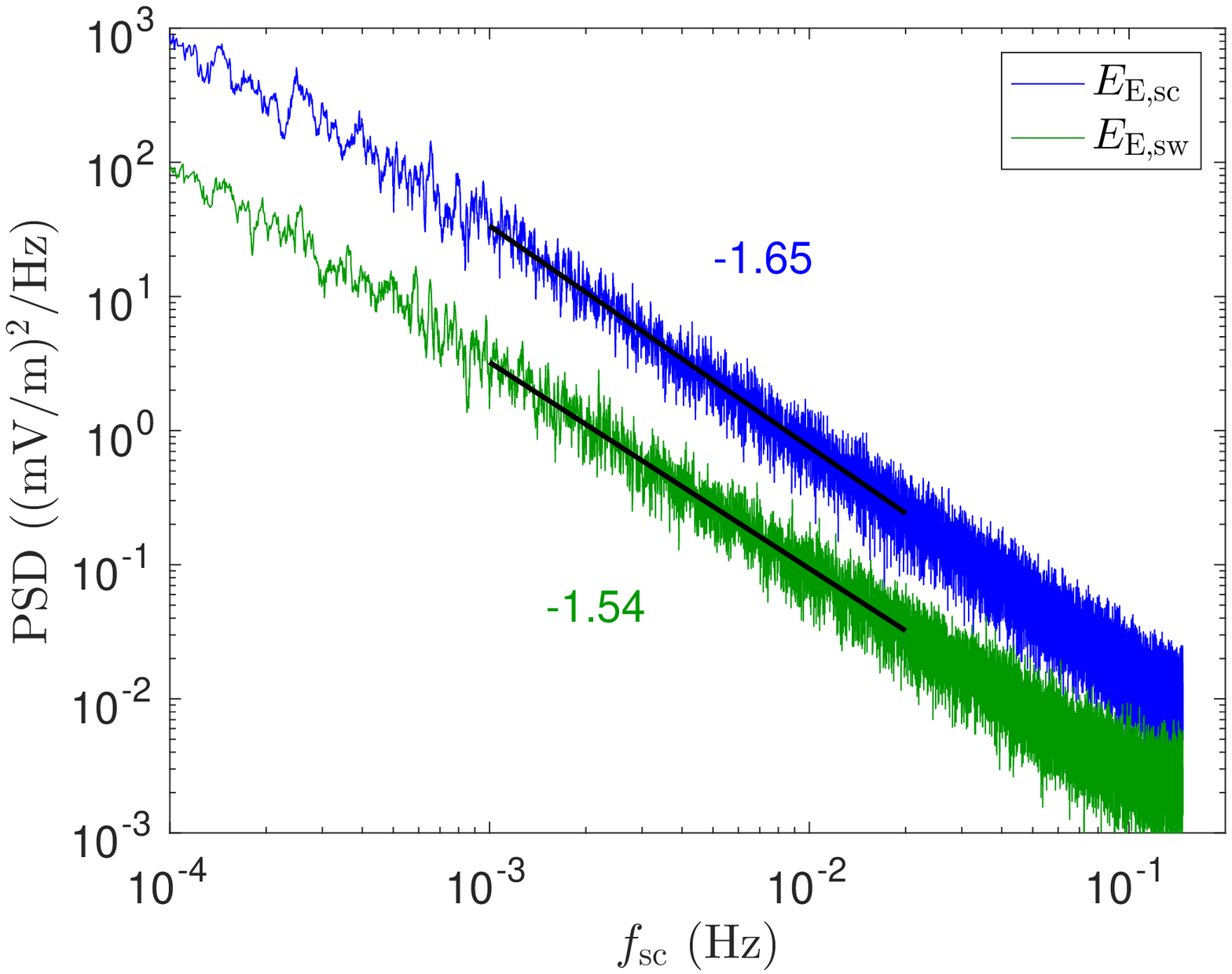}
\hspace{0.04\textwidth}
\includegraphics[width=0.48\textwidth,trim=0 0 0 0,clip]{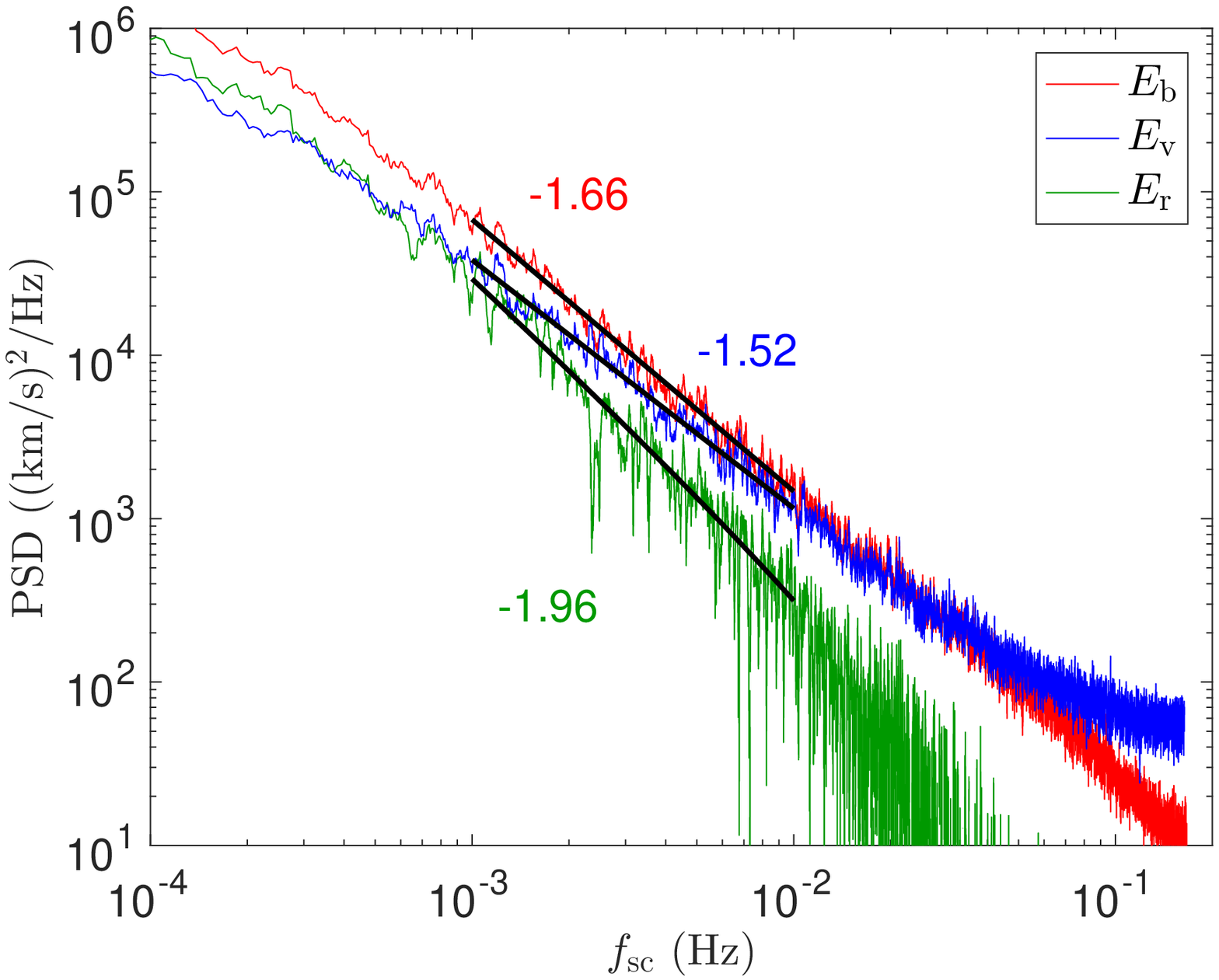}
\caption{Left: electric field spectrum in the spacecraft frame (blue) and plasma frame (green) \citep[adapted from][]{chen11b}. Right: magnetic (red), velocity (blue) and residual energy (green) spectra \citep[adapted from][]{chen13b}. Note that the flattening of $E_\mathrm{v}$ and steepening of $E_\mathrm{r}$ for $f_\mathrm{sc}>10^{-2}$ Hz are artificial (due to instrumental noise).}
\label{fig:spectra}
\end{figure}

While the results of the previous section would appear to favour the \citet{goldreich95} model of Alfv\'enic turbulence, the situation becomes more complicated when also considering the velocity and electric field. The electric field was measured by \citet{bale05} to follow the magnetic field, having a spectral index of $-5/3$ at MHD scales\footnote{Note that because of the anisotropy $k_\perp\gg k_\|$, spectra measured in the solar wind generally correspond to $k_\perp$ spectra (unless the local mean field technique of Section \ref{sec:cb} is used).}. Since the perpendicular plasma velocity in this range should be the $\mathrm{E}\times\mathrm{B}$ drift, the velocity was expected to display a similar spectrum, however, it has been shown to be shallower, with a value nearer $-3/2$ \citep[e.g.,][]{mangeney01,podesta07a}. How can these spectra be consistent with each other? The answer lies in the frame dependence of the electric field \citep{chen11b}. Figure \ref{fig:spectra} shows the electric field spectrum measured in the spacecraft frame $\mathbf{E}_\mathrm{sc}$ with a spectral index near $-5/3$, similar to the magnetic field. It also shows the electric field spectrum after Lorentz transforming to the frame of the mean solar wind velocity $\mathbf{v}_0$,
\begin{equation}
\mathbf{E}_\mathrm{sw}=\mathbf{E}_\mathrm{sc}+\mathbf{v}_0\times\mathbf{B}.
\end{equation}
It can be seen that the solar wind frame spectrum is lower by an order of magnitude and is shallower, having a spectral index close to $-3/2$. These spectra are, therefore, in fact consistent with the velocity and magnetic field measurements, and are independent confirmation of the difference between them. The $\delta\mathbf{E}_\mathrm{sw}$ spectrum matches the velocity spectrum since to leading order it is given by $\delta\mathbf{E}_\mathrm{sw}=-\delta\mathbf{v}\times\mathbf{B}_0$, and the $\delta\mathbf{E}_\mathrm{sc}$ spectrum matches the magnetic spectrum since to leading order it is $\delta\mathbf{E}_\mathrm{sc}=-\mathbf{v}_0\times\delta\mathbf{B}$, i.e., it is dominated by the magnetic fluctuations convected at the mean solar wind velocity.

While the electric, magnetic, and velocity spectra are self-consistent, the question remains why the velocity and magnetic fluctuations differ. This difference is known as the residual energy, and its spectrum is defined as $E_\mathrm{r}(\mathbf{k})=E_\mathrm{v}(\mathbf{k})-E_\mathrm{b}(\mathbf{k})$. A measure of the amount of residual energy is the Alfv\'en ratio $r_\mathrm{A}=\delta\mathbf{v}^2/\delta\mathbf{b}^2$, which is $r_\mathrm{A}=1$ for a pure Alfv\'en wave, but is measured to be $r_\mathrm{A}\approx 0.7$ in the solar wind \citep{chen13b}. Statistical arguments have been made to explain this dominance of magnetic fluctuation energy \citep[e.g.,][]{frisch75,pouquet76,boldyrev12c}. The existence of residual energy, both theoretically and observationally, indicates that strong turbulence can produce quantitative differences to the linear wave relationships.

The residual energy has also been described as a balance between the Alfv\'en effect (linear term) leading to equipartition and a turbulent dynamo (nonlinear term) generating the magnetic excess \citep{grappin83,muller05}. \citet{muller05} used an isotropic closure theory to suggest that $E_\mathrm{r}$ varies with the total energy spectrum $E_\mathrm{t}$, following $k^{-2}$ for a $k^{-3/2}$ total energy spectrum and $k^{-7/3}$ for a $k^{-5/3}$ total energy spectrum. \citet{boldyrev12a} extended this to an anisotropic model in which the perpendicular spectra are $E_\mathrm{r}(k_\perp)\propto k_\perp^{-1}$ for weak and $E_\mathrm{r}(k_\perp)\propto k_\perp^{-2}$ for strong turbulence; in both cases, $E_\mathrm{r}(k_\|,k_\perp)$ is concentrated near the $k_\perp$ axis, similar to the total energy. In the solar wind, the residual energy has an average spectral index of $-1.9$\footnote{While $E_\mathrm{b}$, $E_\mathrm{v}$, $E_\mathrm{t}$, and $E_\mathrm{r}$ cannot all be true power laws, it is not possible to tell which ones are from observations due to the limited scaling range. In the theoretical models, it is $E_\mathrm{t}$ and $E_\mathrm{r}$ which are the power laws, leading to a steeper $E_\mathrm{b}$ and shallower $E_\mathrm{v}$.} \citep{chen13b}, and an example spectrum is shown in Figure \ref{fig:spectra}, along with the velocity and magnetic spectra\footnote{The Alfv\'en unit normalisation for the magnetic field here uses the mean density $\rho_0$, which is appropriate for $\delta\rho/\rho_0\ll1$, as is typically the case in the solar wind (see Section \ref{sec:compressive}).}. There is also evidence that it is anisotropic and concentrated in the region near $k_\|\approx 0$ \citep{yan16}. These observations would be most consistent with the strong turbulence model of \citet{boldyrev12a}, which provides a possible explanation for the different scaling of the velocity and magnetic fluctuations.

\subsection{Three-dimensional anisotropy}
\label{sec:3d}

\begin{figure}
\centering
\includegraphics[scale=0.37]{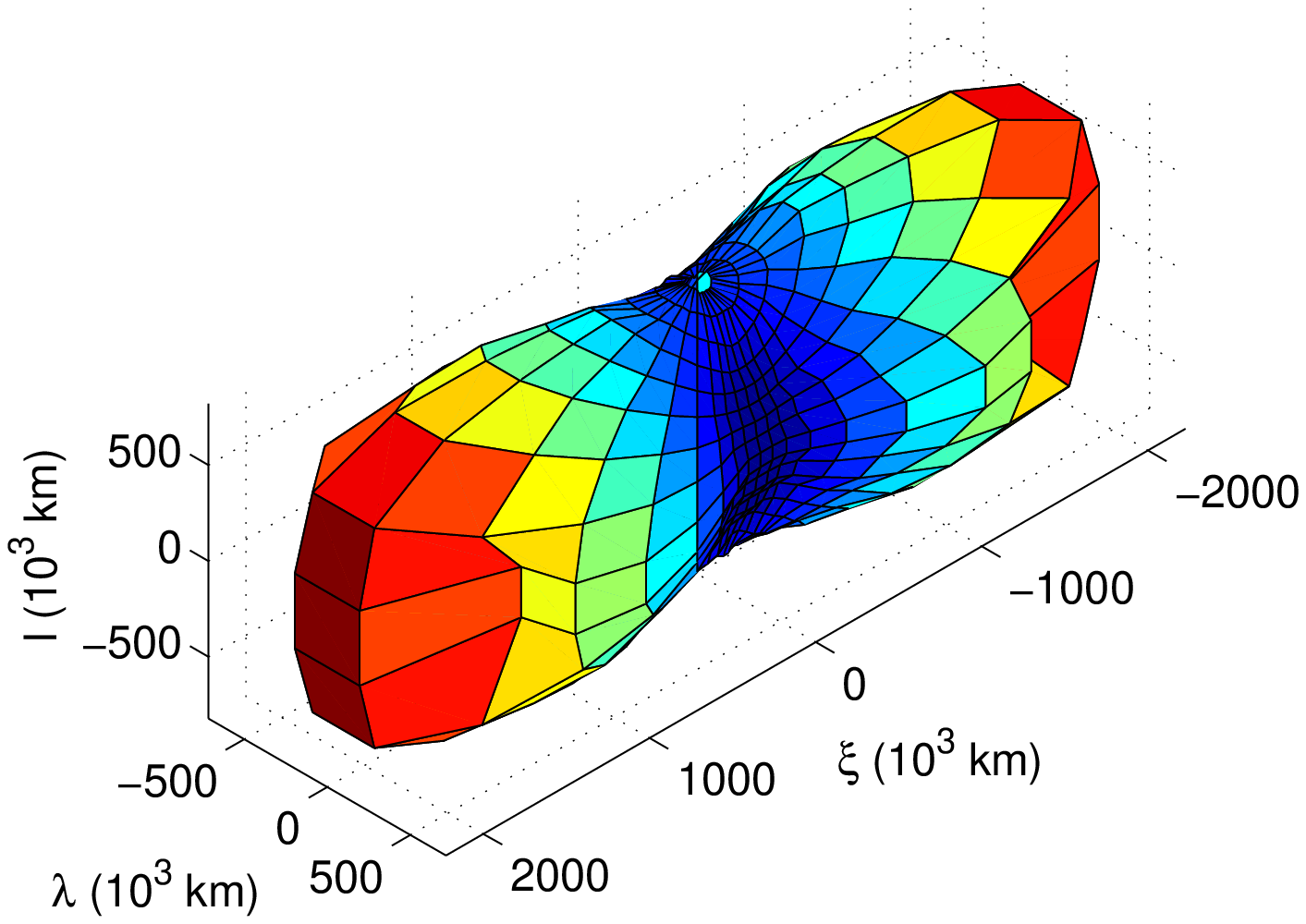}
\includegraphics[scale=0.37]{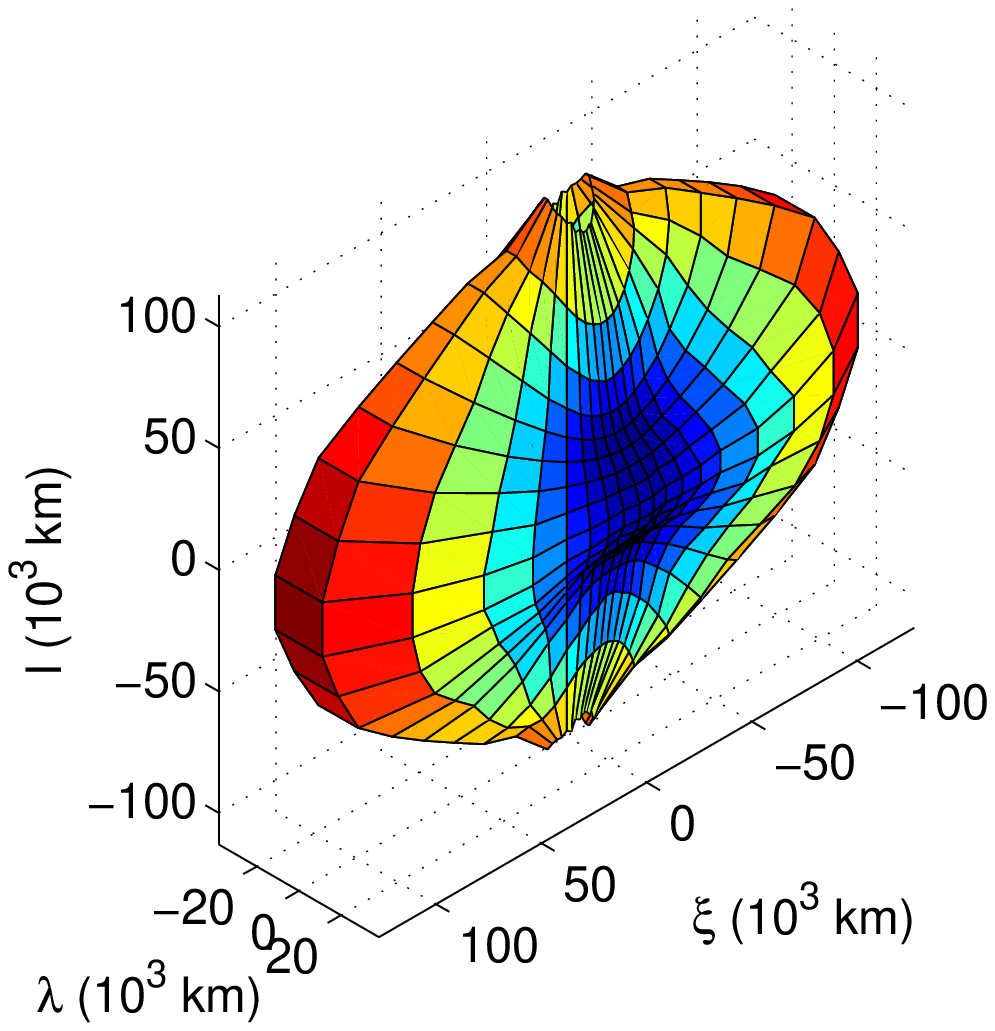}
\includegraphics[scale=0.37]{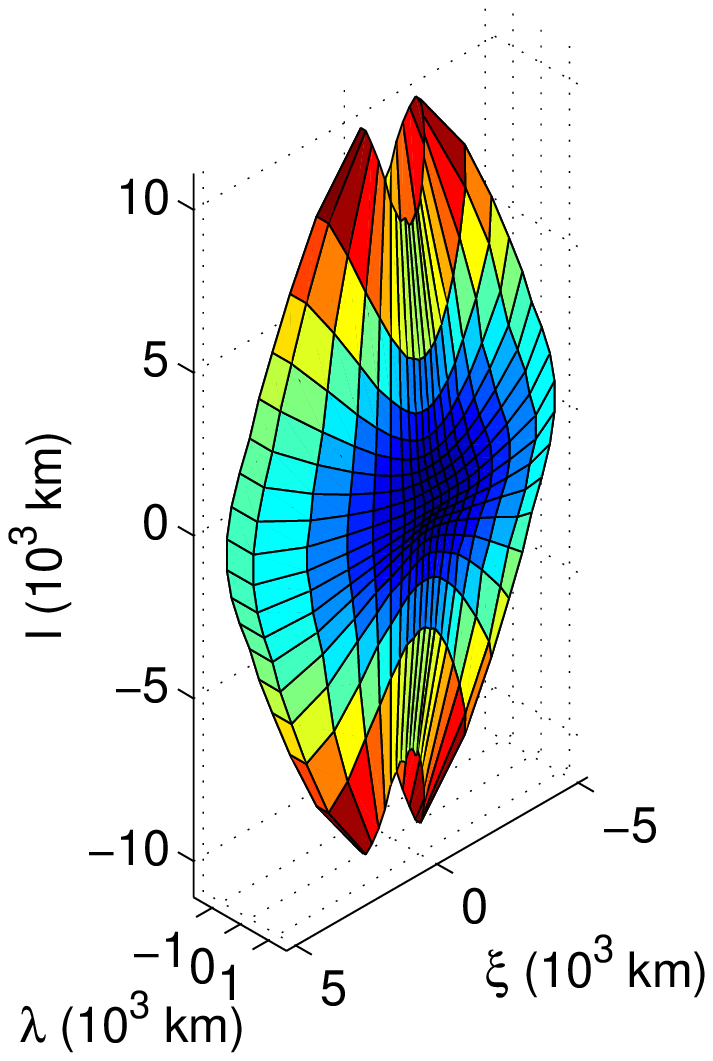}
\caption{3D magnetic eddy shapes from large (left) to small (right) scales, in which $l$ is in the local mean field direction, $\xi$ the local $\delta\mathbf{B}_\perp$ direction, $\lambda$ perpendicular to these, and colour represents distance from the origin \citep[from][]{chen12b}.}
\label{fig:3d}
\end{figure}

Alfv\'enic turbulence is polarised such that magnetic fluctuations are in the plane perpendicular to the mean field, i.e., $\delta B_\perp\gg\delta B_\|$. The local $\delta\mathbf{B}_\perp$ direction within this perpendicular plane breaks the symmetry about the mean field, meaning that the turbulence can be three-dimensionally (3D) anisotropic, e.g., as in the \citet{boldyrev06} model (Section \ref{sec:models}). The techniques of Section \ref{sec:cb} were recently extended to measure this 3D anisotropy in the solar wind \citep{chen12b}. As well as the local mean field, a local $\delta\mathbf{B}$ direction can be defined and a 3D coordinate system constructed, with one axis in the local mean field direction, one in the $\delta\mathbf{B}_\perp$ direction, and the third perpendicular to these. By selecting different $\delta B$ values, and finding the corresponding lengths in each direction, \citet{chen12b} constructed the 3D eddy shapes of the magnetic fluctuations at several scales, shown in Figure \ref{fig:3d}. They vary from being extended in the $\delta\mathbf{B}_\perp$ direction at large scales, to being 3D anisotropic in the sense of \citet{boldyrev06} at small scales. While the small scale anisotropy ($l_\|>\xi>\lambda$) and the scaling in the $l_\|$ and $\xi$ directions is consistent with the \citet{boldyrev06} model, the $\lambda$ scaling is not\footnote{Another test of this model is the scaling of the alignment angles, however, this is difficult to measure due to current instrumental limitations \citep{podesta09e,wicks13b}.}.

There are a few considerations when interpreting these results. Firstly, one might expect the anisotropy with respect to the $\delta\mathbf{B}$ direction to be a simple result of the solenoidality of the magnetic field. However, it can be shown that solenoidality alone does not determine the anisotropy in the local perpendicular plane \citep{chen12b,mallet16}, and extra physics, such as alignment \citep{boldyrev06} is involved. Secondly, it has been suggested that the anisotropy in the perpendicular plane at large scales could be a result of the solar wind expansion as it travels away from the Sun. \citet{verdini15a} argued that magnetic flux conservation in the expanding wind causes fluctuations in the radial direction to be smaller, leading to a bias towards lower fluctuation amplitudes in the $\xi$ direction when measurements are made in the radial direction. Some evidence in support of this hypothesis has recently been reported \citep{vech16}. As well as the eddy shapes, this effect could also alter the 3D scaling, since any effect of the expansion relative to the turbulent dynamics is likely to be scale-dependent. Finally, as discussed in Section \ref{sec:indices}, the situation is further complicated by the presence of residual energy, which causes the magnetic spectrum to be steeper than the velocity spectrum; how this impacts the 3D anisotropy is not yet well understood. Therefore, while observations show that Alfv\'enic turbulence in the solar wind is locally three-dimensionally anisotropic, more work is required to fully understand this.

\subsection{Imbalanced turbulence}

As shown in Section \ref{sec:indices}, the scaling of the velocity and magnetic fluctuations is different, so it is the total energy spectrum $E_\mathrm{t}(\mathbf{k})=E_\mathrm{v}(\mathbf{k})+E_\mathrm{b}(\mathbf{k})$ which should be compared to the models of Section \ref{sec:models}. These models, however, describe balanced Alfv\'enic turbulence, i.e., turbulence with equal fluxes of Alfv\'enic fluctuations propagating in each direction along the mean field. In real systems, turbulence is often imbalanced, with unequal fluxes created by localised driving sources, such as supernovae in the interstellar medium or Alfv\'en waves from the Sun.  In the solar wind, the level of imbalance can be quantified with the normalised cross helicity $\sigma_\mathrm{c}=2\left<\delta\mathbf{v}\cdot\delta\mathbf{b}\right>/\left<\delta\mathbf{v}^2+\delta\mathbf{b}^2\right>$; $\sigma_\mathrm{c}\approx 0$ corresponds to balanced turbulence and $\sigma_\mathrm{c}\approx\pm 1$ to highly imbalanced turbulence. While the solar wind at 1 AU has a typical imbalance of $\sigma_\mathrm{c}=0.46$ \citep{chen13b}, there is significant variability, enabling the systematic dependence to be studied.

\begin{figure}
\centering\includegraphics[width=0.6\textwidth,trim=0 0 0 0,clip]{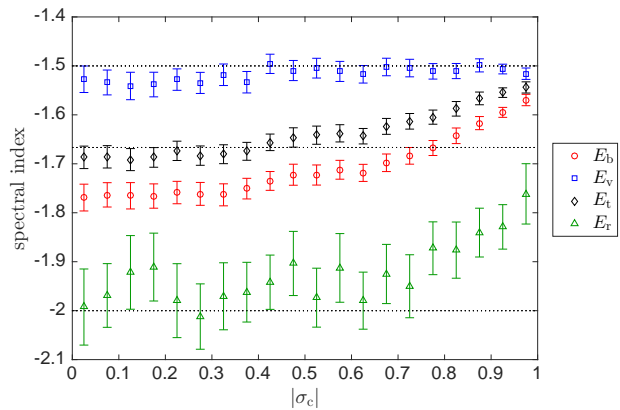}
\caption{Variation of spectral indices of magnetic field ($E_\mathrm{b}$), velocity ($E_\mathrm{v}$), total energy ($E_\mathrm{t}$), and residual energy ($E_\mathrm{r}$) with the level of imbalance $|\sigma_\mathrm{c}|$ \citep[adapted from][]{chen13b}.}
\label{fig:sigmac}
\end{figure}

Figure \ref{fig:sigmac} shows the dependence of the spectral indices on the level of imbalance (all quantities are measured in the middle of the inertial range). The total energy spectral index varies from $\approx -5/3$ at low imbalance to $\approx -3/2$ at high imbalance \citep[as also found by][]{podesta10d}.  $E_\mathrm{b}$ and $E_\mathrm{v}$ scale similarly for $|\sigma_\mathrm{c}|\approx 1$, which is consistent with the residual energy being mathematically constrained to be small for large imbalance. It is interesting, however, that they take a value close to $-3/2$, favouring the \citet{boldyrev06} model in the absence of residual energy. The residual energy itself has a spectral index of $-2$ (see Section \ref{sec:indices}) for low imbalance, which becomes shallower for larger imbalance\footnote{Although as noted by \citet{chen13b}, because the amount of residual energy is small at large imbalance, there may be a systematic bias towards shallower spectra here due to small inaccuracies in the magnetic field normalisation.}. Both Elsasser variables (not shown here) have the same scaling as the total energy, to within uncertainties, for all $|\sigma_\mathrm{c}|$\footnote{This is true except for the two largest $|\sigma_\mathrm{c}|$ bins where the $\mathbf{z}^-$ spectrum is affected by instrumental noise.}. The main dependence of the total energy spectrum with $\sigma_\mathrm{c}$, however, is not predicted by any of the current models of imbalanced Alfv\'enic turbulence \citep[e.g.,][]{lithwick07,beresnyak08,chandran08,perez09,podesta10c}. Therefore, while several aspects of Alfv\'enic turbulence in the solar wind are beginning to be understood, a complete explanation still remains to be found.

\section{Compressive fluctuations}
\label{sec:compressive}

\begin{figure}
\centering
\includegraphics[width=0.45\textwidth,trim=0 0 0 0,clip]{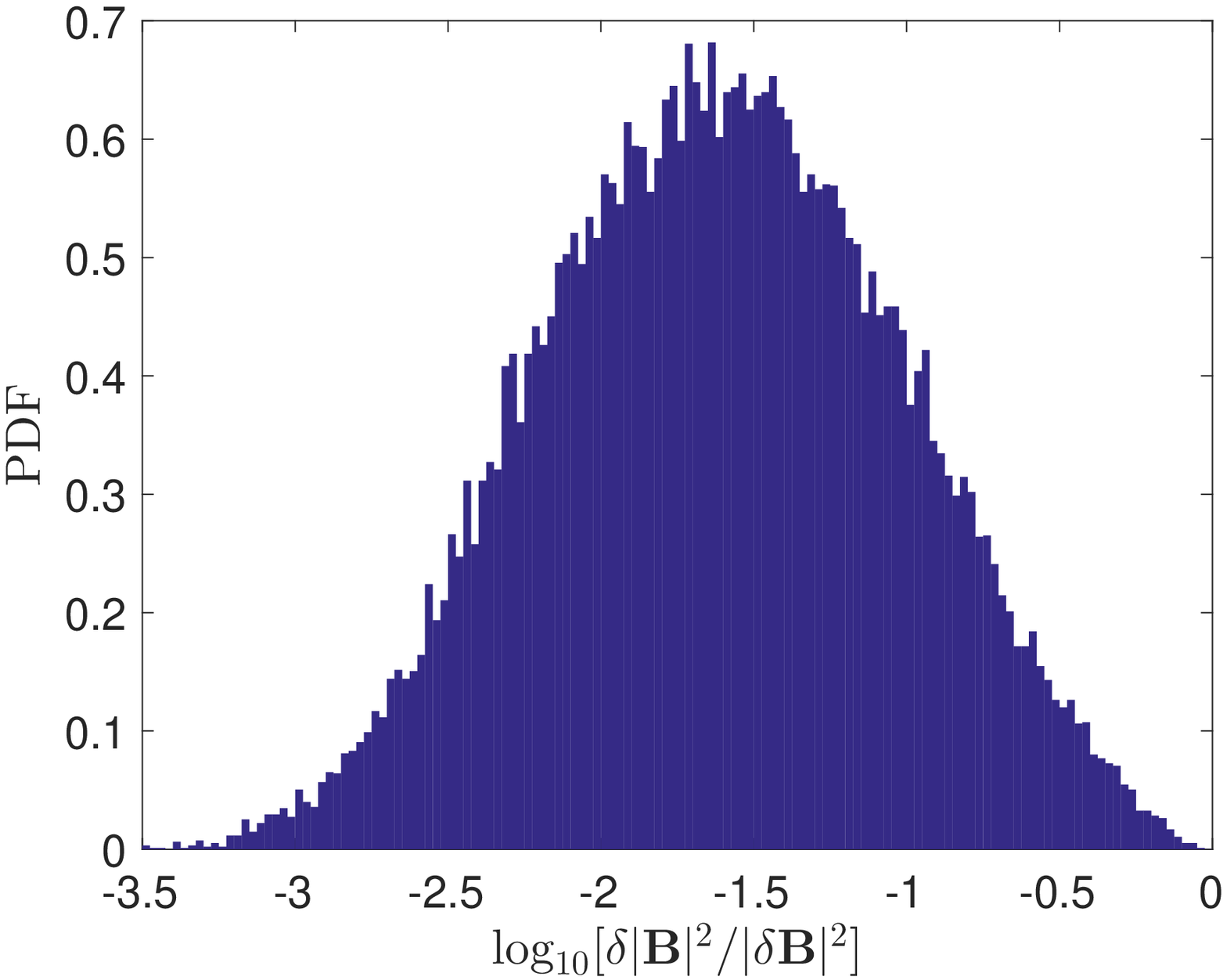}
\hspace{0.04\textwidth}
\includegraphics[width=0.45\textwidth,trim=0 0 0 0,clip]{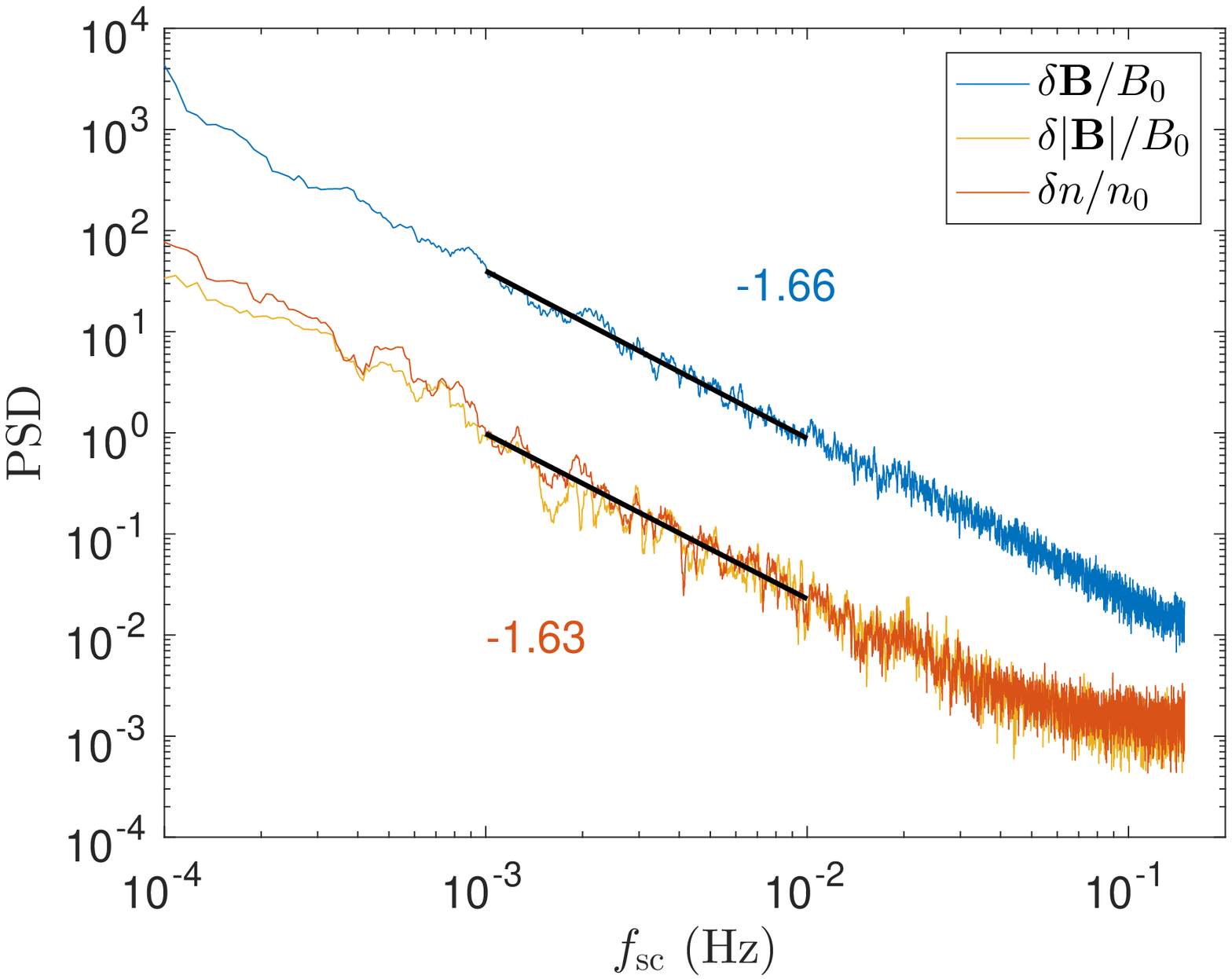}
\caption{Left: distribution of magnetic compressibility at the outer scale of the Alfv\'enic inertial range. Right: normalised spectra of compressive fluctuations, $\delta n$ and $\delta |\mathbf{B}|$, in comparison to the total magnetic fluctuation spectrum, $\delta\mathbf{B}$.}
\label{fig:comphistspectra}
\end{figure}

While the dominant power in the solar wind is in the Alfv\'enically polarised fluctuations, there is also a measurable fraction in non-Alfv\'enic modes, involving variations of the density $\delta n$ and magnetic field magnitude $\delta|\mathbf{B}|$. As well as being of intrinsic interest, understanding this compressive component is particularly important for interpreting measurements of plasmas outside our solar system, where density fluctuations are usually the most observationally accessible quantity, for example in the interstellar medium \citep{armstrong95} and galaxy clusters \citep{zhuravleva14short}. They are also thought to play a key role in the star formation process \citep{mckee07}. The distribution of the magnetic compressibility in the solar wind at the 1 hour scale ($k\rho_\mathrm{i}\sim 10^{-3}$) is shown in Figure \ref{fig:comphistspectra}; the average fraction of magnetic energy in the compressive component is $\sim$2\%\footnote{When using $\delta B_\|$ instead of $\delta|\mathbf{B}|$ to define the magnetic compresibility this value is $\sim$10\%.}. What is the nature of this component and how does it interact with the Alfv\'enic turbulence? In this Section, our recent progress on these questions is discussed.

\subsection{Spectra and passivity}

It was proposed by \citet{higdon84} that the density variations observed in interstellar turbulence are entropy fluctuations, and possibly magnetosonic waves, passively mixed by the Alfv\'enic turbulence. This concept was developed further by \citet{goldreich97}, who suggested that for MHD turbulence, slow waves (or pseudo-Alfv\'en waves in the incompressible limit) would be passive to the Alfv\'enic turbulence as a result of the anisotropy $k_\perp\gg k_\|$ developed by the cascade. An Alfv\'en wave ($\delta\mathbf{z}_\mathrm{A}^\pm$) is polarised perpendicular to $\mathbf{B}_0$, and distorts a counter-propagating Alfv\'en wave packet through the term $\delta\mathbf{z}_\mathrm{A}^\pm\cdot\nabla\delta\mathbf{z}_\mathrm{A}^\mp$, which is of order $\sim k_\perp\delta z_\mathrm{A}^2$. However, an oblique slow wave ($\delta\mathbf{z}_\mathrm{s}^\pm$) is polarised mostly parallel to $\mathbf{B}_0$, so distorts a counter-propagating Alfv\'en wave packet through the term $\delta\mathbf{z}_\mathrm{s}^\pm\cdot\nabla\delta\mathbf{z}_\mathrm{A}^\mp$, which is of order $\sim\delta z_\mathrm{s} k_\|\delta z_\mathrm{A}$, i.e., a factor of $k_\|/k_\perp$ smaller (assuming equal amplitudes). This would mean that slow wave fluctuations are scattered by the anisotropic Alfv\'enic turbulence, but do not actively interfere with it, which would also result in negligible energy transfer from the Alfv\'enic to slow mode component \citep{goldreich95,maron01}. \citet{schekochihin09} argued that for weakly collisional plasmas such as the solar wind, the MHD description is not sufficient for the compressive fluctuations\footnote{Although for the Alfv\'enic component that it would be sufficient, since small amplitude ($\delta B/B_0\ll1$) Alfv\'en waves at $k_\perp\rho_\mathrm{i}\ll1$ do not modify the Maxwellian nature of the ion distribution \citep{schekochihin09}.}, and a kinetic treatment is required. In the gyrokinetic formalism, where $k_\perp\gg k_\|$ is taken as an ordering assumption, the nonlinear fluid equations for the Alfv\'enic turbulence at $k_\perp\rho_\mathrm{i}\ll1$ (Reduced MHD) decouple from the ion kinetic equation. The active Alfv\'enic turbulence is self-contained, and passively mixes the non-Alfv\'enic part of the ion distribution, from which the compressive fluctuations (which produce the kinetic counterparts of the slow and entropy modes) are obtained \citep{schekochihin09}. Therefore, the passivity of these modes, originally derived in the context of fluid models, is predicted to hold for kinetic plasmas too.

In hydrodynamic turbulence, a passive scalar $\sigma$ shares the same spectrum as the advecting velocity field $\mathbf{v}$ (neglecting intermittency). This is because it follows its continuity equation, $\partial\sigma/\partial t+\mathbf{v}\cdot\nabla\sigma=0$, meaning that its nonlinear time $\sim l/\delta v$ is the same as that of the active turbulence, leading to $\sigma\propto\delta v$. In plasma turbulence, however, the situation is more complicated, since the nonlinear times involve both the velocity and magnetic fields, and their alignment angles (see Section \ref{sec:models}). It is therefore of interest to compare the scaling of the compressive fluctuations in the solar wind to that of the Alfv\'enic turbulence. While spectra of compressive fluctuations have long been measured in the solar wind \citep[see reviews by][]{bruno13,alexandrova13a}, we have recently been able to investigate these with greater precision, to enable comparison to the Alfv\'enic turbulence. \citet{chen11b} showed that the spectral indices of both $\delta n$ and $\delta |\mathbf{B}|$ are close to $-5/3$, similar to the magnetic field, rather than the velocity, which has a $-3/2$ spectral index. An example from this data set is shown in Figure \ref{fig:comphistspectra}, where the typical features can be seen: compressive fluctuations at a level much lower than the total power, with a spectral slope similar to that of the Alfv\'enic magnetic fluctuations. This scaling of the compressive fluctuations indicates that they are not passively advected in the hydrodynamic sense. It is still possible, however, that they are passive to the Alfv\'enic turbulence, in which they are mixed by both the velocity and magnetic fluctuations.

\subsection{Nature of the compressive fluctuations}

\begin{figure}
\centering
\includegraphics[width=0.6\textwidth,trim=0 0 0 0,clip]{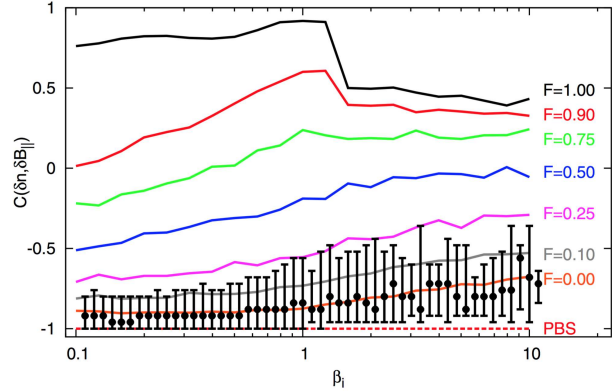}
\caption{Correlation $C$ of density and parallel magnetic fluctuations as a function of ion plasma beta $\beta_\mathrm{i}$, compared to theoretical predictions for a spectrum of kinetic fast and slow mode waves with different fractions $F$ of fast mode to total energy \citep[from][]{howes12a}.}
\label{fig:howes12}
\end{figure}

It is also of interest to determine what type of compressive fluctuations are present, i.e., which wave modes they resemble, and whether these require a kinetic description. \citet{klein12} showed that the kinetic fast and slow modes retain qualitative characteristics of their MHD counterparts, but differ quantitatively so that a kinetic description is necessary. In particular, $\delta n$ and $\delta B_\|$ remain predominantly anti-correlated for the slow mode and correlated for the fast mode, with a characteristic ion beta, $\beta_\mathrm{i}$, dependence. \citet{howes12a} compared solar wind measurements of this correlation to that produced from a critically balanced spectrum of kinetic slow waves plus an isotropic spectrum of kinetic fast waves, with different fractions of the two. The results are given in Figure \ref{fig:howes12} and show a strong, $\beta_\mathrm{i}$-dependent anti-correlation, consistent with the curve in which only slow modes, and not fast modes, contribute to the compressive power. The scarcity of fast mode fluctuations has some interesting implications. Firstly, it helps justify the application of low frequency approximations, such as gyrokinetics \citep{schekochihin09} and kinetic reduced MHD \citep{kunz15} to the solar wind. Secondly, it suggests that there can be little transfer of energy to whistler turbulence at sub-ion scales (consistent with observations; Section \ref{sec:nature}), which in turn constrains the possible dissipation mechanisms (Section \ref{sec:dissipation}).

\subsection{Anisotropy and damping}

An interesting question is why there appears to be a cascade of compressive fluctuations at all, when they are expected to be strongly damped in a $\beta_\mathrm{i}\sim 1$ plasma \citep{barnes66}. In particular, their damping rate is $\gamma\sim k_\|v_\mathrm{A}$, comparable to the timescale on which the Alfv\'enic turbulence would cascade them to small scales. The solution to this may lie in their anisotropy: if they are significantly more anisotropic than the Alfv\'enic turbulence, i.e., their $k_\|$ remains small, they are not strongly damped. So what do we expect the anisotropy of passively mixed compressive fluctuations to be? \citet{lithwick01} suggested that the slow modes inherit the anisotropy of the Alfv\'enic turbulence that mixes them. However, \citet{schekochihin09} argued that the kinetic equation for the compressive fluctuations becomes linear in the Lagrangian frame of the Alfv\'enic turbulence, and, therefore, that they have no parallel cascade. This would lead to highly elongated, compressive structures, with very small $k_\|$, and could explain why they are not strongly damped.

\begin{figure}
\centering
\includegraphics[width=0.3\textwidth,trim=0 0 0 0,clip]{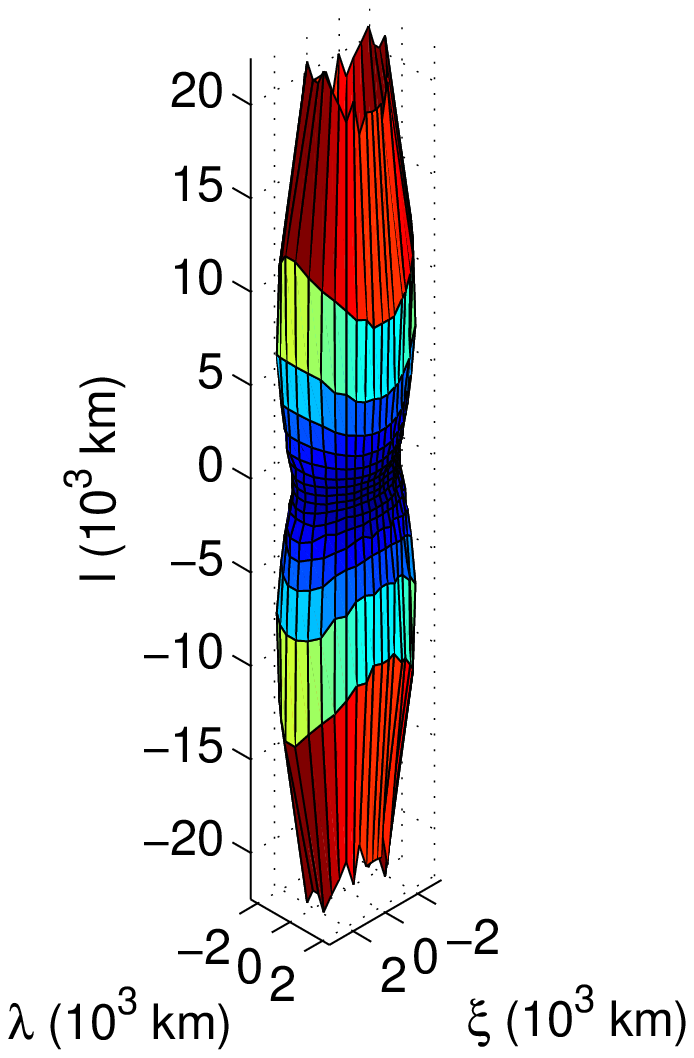}
\hspace{0.04\textwidth}
\includegraphics[width=0.6\textwidth,trim=0 0 0 0,clip]{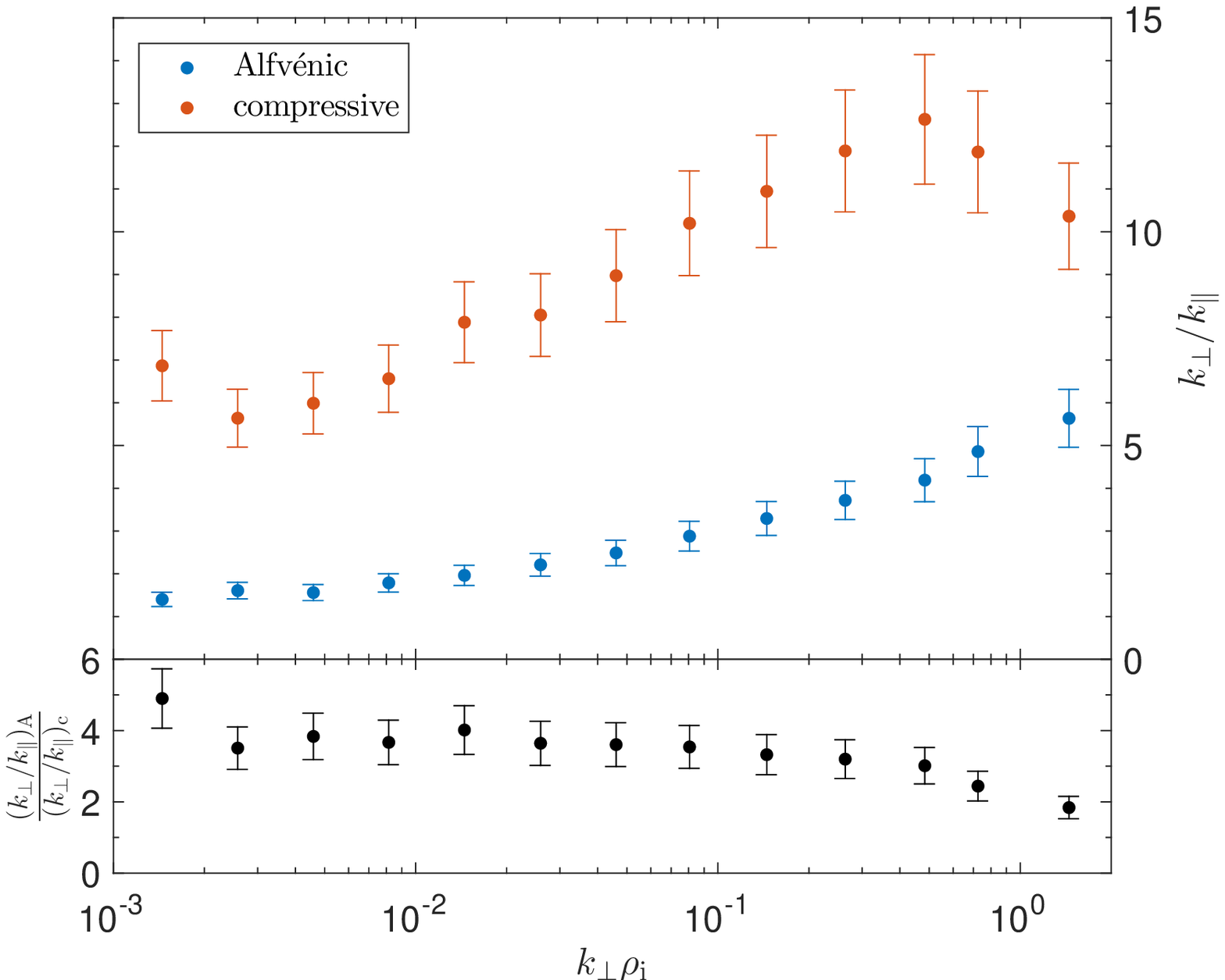}
\caption{Left: 3D eddy shape for the $\delta|\mathbf{B}|$ component of the turbulence at small scales ($k_\perp\rho_\mathrm{i}\approx 0.4$) in the same form as Figure \ref{fig:3d} \citep[from][]{chen12b}. Right: comparison between the anisotropy of the Alfv\'enic and compressive components of the turbulence.}
\label{fig:compressiveanisotropy}
\end{figure}

The 3D shape of the $\delta |\mathbf{B}|$ fluctuations was measured (using the technique described in Section \ref{sec:3d}) by \citet{chen12b} and is shown in Figure \ref{fig:compressiveanisotropy}, where it can be seen that they are indeed very anisotropic. They are, in fact, more anisotropic than the Alfv\'enic turbulence, as illustrated in Figure \ref{fig:compressiveanisotropy} where the anisotropy $k_\perp/k_\|$ of each component is shown as a function of scale. The ratio of the anisotropy of the compressive to Alfv\'enic component is shown in the lower panel and can be seen to take a value around 4 throughout most of the inertial range\footnote{At $k_\perp\rho_\mathrm{i}\sim 1$ the compressive component anisotropy reduces, becoming closer to that of the Alfv\'enic component, consistent with the $\delta B_\|$ of the  Alfv\'enic turbulence starting to dominate the compressive spectrum.}. Due to the angular resolution of the measurements, this is a lower limit: the compressive fluctuations are at least several times more anisotropic than the Alfv\'enic turbulence. This is consistent with the prediction of \citet{schekochihin09}, although there is an alternative interpretation. It is possible that the cascade attempts to generate compressive fluctuations with an anisotropy similar to the Alfv\'enic turbulence, but these are quickly damped, leaving the highly anisotropic ones to be observed. So the question becomes: are the compressive fluctuations highly anisotropic because the less anisotropic component is damped, or are they generated highly anisotropic to begin with? This remains to be answered, but has some interesting implications, e.g., if damping is taking place, then energy may be being removed from the cascade throughout the inertial range, rather than just at the small scales. However, a recent suggestion by \citet{schekochihin16} is that ``anti-phase-mixing'' may effectively suppress such damping, providing an alternative explanation for the presence of the compressive fluctuations. Further observations will be required to distinguish these possibilities.

\section{Kinetic range turbulence}
\label{sec:kineticrange}

As the energy cascade proceeds to ever smaller scales, it eventually reaches plasma microscales such as the particle gyroradii and inertial lengths. Important changes occur here. For a start, scale-invariance is broken and the energy spectrum deviates from its power law form, but it is also in this range where dissipation and heating are thought to take place. Several names have been proposed for this range (and and the various sub-ranges within), reflecting different possible physical processes. The term ``kinetic range'', reflecting the fact that the scales are close to the particle gyroradii, has recently gained popularity and will be used here\footnote{Although this terminology is not ideal, since aspects of the turbulence in the ``kinetic range'' can be captured by fluid models, and aspects of turbulence in the larger-scale (MHD) range require a kinetic treatment (Section \ref{sec:compressive}).}. Since the resolution of remote astrophysical observations is limited, the solar wind presents an unparalleled opportunity to understand kinetic range turbulence and how it leads to plasma heating. In this section, our recent progress on kinetic range turbulence is discussed.

\subsection{Phenomenological models of kinetic range turbulence}
\label{sec:kineticmodels}

Similarly to Alfv\'enic turbulence (Section \ref{sec:models}), phenomenological models have been developed for an energy cascade between ion and electron scales. Early models were based on the fluid equations of Electron MHD (EMHD) \citep[e.g.,][]{kingsep90}, which in fluctuating form are
\begin{equation}
\frac{\partial\delta\mathbf{B}}{\partial t}=\frac{1}{\mu_0en_\mathrm{e}}\nabla\times\left[\mathbf{B}_0\times\left(\nabla\times\delta\mathbf{B}\right)+\delta\mathbf{B}\times\left(\nabla\times\delta\mathbf{B}\right)\right].
\label{eq:emhd}
\end{equation}
This is essentially the induction equation with the ions assumed motionless (and electron inertia and dissipation terms also neglected). While this turns out to be a poor assumption for the solar wind kinetic range (see Section \ref{sec:nature}), the form of the EMHD equations can be used to illustrate some of the basic principles of the kinetic range cascade.

\begin{figure}
\centering
\includegraphics[width=0.5\textwidth,trim=0 0 0 0,clip]{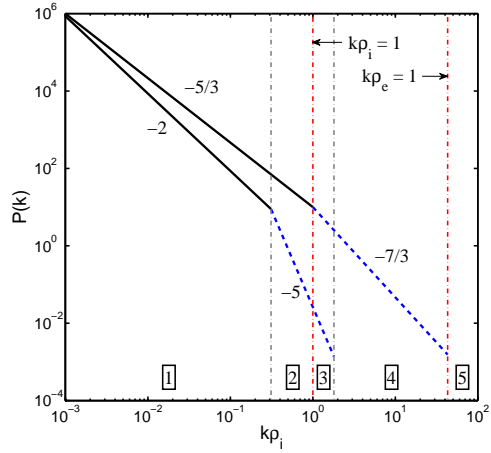}
\caption{Schematic of parallel and perpendicular energy spectra for critically balanced Alfv\'enic turbulence ($k_\perp^{-5/3}$ and $k_\|^{-2}$) at $k_\perp\rho_\mathrm{i}<1$, and kinetic Alfv\'en or whistler turbulence ($k_\perp^{-7/3}$ and $k_\|^{-5}$) at $k_\perp\rho_\mathrm{i}>1$, without intermittency or other corrections \citep[from][]{chen10a}.}
\label{fig:chen10a}
\end{figure}

Equation \ref{eq:emhd} is similar to Equation \ref{eq:mhd}: there is a linear term (describing whistler waves) and a nonlinear term responsible for the turbulence. The key difference, however, is the additional spatial gradient in both terms. This results in whistler waves being dispersive and leads to a steeper turbulence spectrum. This was first shown by \citet{vainshtein73}, who considered isotropic strong EMHD turbulence. From the form of the nonlinear term in Equation \ref{eq:emhd} it can be seen that the nonlinear time is $\tau_\mathrm{nl}\propto l^2/\delta B$. Assuming a constant energy cascade rate $\varepsilon\propto \delta B^2/\tau_\mathrm{nl}\propto\delta B^3/l^2$ leads to a magnetic energy spectrum
\begin{equation}
E_\mathrm{B}(k)\propto k^{-7/3},
\end{equation}
steeper than any of the predicted spectra for MHD turbulence. Similarly to MHD turbulence, though, the assumption of isotropy is not robust. When a critical balance between the whistler timescale and nonlinear timescale is assumed, the scale-dependent anisotropy $k_\|\propto k_\perp^{1/3}$ is obtained \citep{cho04}, a stronger anisotropy than for Alfv\'enic turbulence. This results in anisotropic spectra
\begin{equation}
E_\mathrm{B}(k_\perp) \propto k_\perp^{-7/3},\ \ \ E_\mathrm{B}(k_\|) \propto k_\|^{-5},
\end{equation}
as illustrated in Figure \ref{fig:chen10a}. 

It was proposed \citep[e.g.,][]{leamon98a,howes08a,schekochihin09,boldyrev12b}, however, that Alfv\'enic turbulence at large scales transitions to kinetic Alf\'ven turbulence at sub-ion scales, rather than whistler/EMHD turbulence, reflecting the fact that the kinetic Alfv\'en wave (KAW) is the continuation of the oblique Alfv\'en mode in this regime. The nonlinear fluid equations that capture kinetic Alfv\'en turbulence between the ion and electron gyroscales $1/\rho_\mathrm{i}\ll k_\perp\ll 1/\rho_\mathrm{e}$ \citep[e.g.,][]{schekochihin09,boldyrev13a} take the same mathematical structure as the EMHD equations for a strong mean field and anisotropy $k_\perp\gg k_\|$ \citep{boldyrev13a}, so the above scaling predictions for the magnetic fluctuations also apply. A key physical difference, however, is that kinetic Alfv\'en turbulence is low frequency, $\omega\ll k_{\perp}v_\mathrm{th,i}$, (whereas whistler turbulence has $\omega\gg k_{\perp}v_\mathrm{th,i}$) so unlike whistler turbulence, density fluctuations are non-negligible \citep{chen13c,boldyrev13a}. From the nonlinear kinetic Alfv\'en equations, the scaling of the magnetic, electric and density fluctuations is
\begin{equation}
E_\mathrm{B}(k_\perp) \propto k_\perp^{-7/3},\ \ \ E_\mathrm{E}(k_\perp) \propto k_\perp^{-1/3},\ \ \ E_\mathrm{n}(k_\perp) \propto k_\perp^{-7/3}.
\label{eq:kawspectrum}
\end{equation}

A modification of these predictions was suggested by \citet{boldyrev12b}, who observed strong intermittency in their kinetic Alfv\'en turbulence simulations, with fluctuation energy concentrated in 2D sheets. Assuming that the cascade is only active within these sheets (so that the active volume fraction is proportional to the scale of the fluctuations), a weaker anisotropy is obtained, $k_\|\propto k_\perp^{2/3}$ and the total energy spectra are
\begin{equation}
E(k_\perp) \propto k_\perp^{-8/3},\ \ \ E(k_\|) \propto k_\|^{-7/2}.
\end{equation}
As for Alfv\'enic turbulence, these predictions for the spectra, nature, and intermittency of the fluctuations can be directly tested with solar wind observations.

\subsection{Kinetic range spectra}

\begin{figure}
\centering
\includegraphics[width=0.55\textwidth,trim=0 0 0 0,clip]{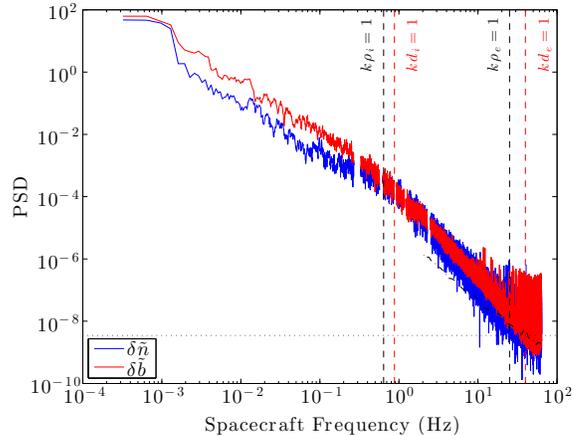}
\caption{Spectra of density and magnetic fluctuations normalised according to Equation \ref{eq:kawnorm}; the vertical dashed lines correspond to the ion and electron gyroradii and inertial length scales under the assumption of the Taylor hypothesis \citep[from][]{chen13b}.}
\label{fig:nbspectra}
\end{figure}

It has been known for a long time that the spectrum of magnetic fluctuations in the solar wind steepens at ion scales \citep[e.g.,][]{coleman68,russell72}, but it is only recently that we have been able to more comprehensively diagnose this range, using new high resolution spacecraft instrumentation. While the shape of the magnetic spectrum close to ion scales and close to electron scales is somewhat variable, the range in between is generally found to be a power law with a typical spectral index around $-2.8$ \citep[e.g.,][]{alexandrova09,sahraoui13a}. More recently, we have also been able to measure the density spectrum in this range \citep{chen12a,safrankova13a,safrankova15}. \citet{chen12a} showed that this also takes a power law form, with a spectral index of $-2.75\pm0.06$, similar to the magnetic spectrum. Typical spectra of the density and magnetic fluctuations are shown in Figure \ref{fig:nbspectra}. Electric field spectra have also been reported to flatten at ion scales \citep{bale05,sahraoui09,salem12}, and ion velocity and temperature spectra to steepen \citep{safrankova13a,safrankova16}.

The fact that the density and magnetic fluctuations have the same scaling as each other is consistent with expectations for kinetic Alfv\'en turbulence, in which these two fields are directly coupled \citep{schekochihin09,boldyrev13a}, but the spectral index of $-2.8$ is steeper than the pure cascade prediction of $-7/3$ (Equation \ref{eq:kawspectrum})\footnote{This comparison assumes the Taylor hypothesis, which is thought to be valid for kinetic Alfv\'en turbulence at 1 AU but not whistler turbulence \citep{howes14b,klein14b}.}. Several explanations have been proposed to explain the steeper scaling. For example, \citet{howes11a} also observed a $k_\perp^{-2.8}$ magnetic spectrum in their gyrokinetic simulations, interpreting the steeper-than-$k_\perp^{-7/3}$ result as due to the presence of electron Landau damping. The steeper spectrum, however, has also been observed in simulations that do not contain such damping \citep{boldyrev12b,franci15b}. As mentioned in Section \ref{sec:kineticmodels}, \citet{boldyrev12b} suggested that the accumulation of the cascade into intermittent 2D structures would lead to a $k_\perp^{-8/3}$ spectrum, which is closer to the observed value. Other explanations have also been proposed \citep[e.g.,][]{meyrand13,passot15}. It remains to be determined which combination of these, or other possibilities, are responsible for the steep kinetic range spectra.

\subsection{Nature of the fluctuations}
\label{sec:nature}

\begin{figure}
\centering
\includegraphics[width=0.6\textwidth,trim=0 0 0 0,clip]{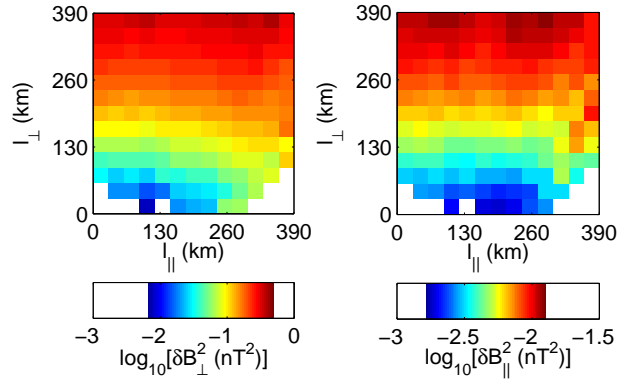}
\caption{Fluctuation power in the perpendicular (left) and parallel (right) magnetic field components as a function of perpendicular ($l_\perp$) and parallel ($l_\|$) lengthscale, for the kinetic range between the ion and electron gyroscales, $1/\rho_\mathrm{i}<k<1/\rho_\mathrm{e}$ \citep[from][]{chen10b}.}
\label{fig:kineticanisotropy}
\end{figure}

To understand how the kinetic scale cascade operates and leads to plasma heating, it is necessary to determine the nature of the fluctuations which constitute the turbulence. As mentioned in Section \ref{sec:alfvenicturbulence}, turbulence at large scales, $k\rho_\mathrm{i}\ll1$, is predominantly Alfv\'enic, i.e., the fluctuations are polarised similarly to Alfv\'en waves \citep{belcher71}. The plasma modes which exist in the kinetic range, however, are more complex. To narrow down the possibilities, it is first helpful to determine the anisotropy of the turbulence, i.e., whether it is dominated by parallel ($k_\|\gg k_\perp$), isotropic ($k_\perp\sim k_\|$) or perpendicular ($k_\perp\gg k_\|$) fluctuations. \citet{chen10b} measured the anisotropy of the magnetic fluctuations in the kinetic range using a technique similar to that described in Section \ref{sec:cb} but with multiple spacecraft to simultaneously measure different directions. The result of this analysis is given in Figure \ref{fig:kineticanisotropy}, which shows the power in the perpendicular and parallel field components as a function of parallel and perpendicular scale. For both components, the power contours are elongated in the parallel direction, indicating that the turbulence consists of perpendicular fluctuations, $k_\perp\gg k_\|$. This anisotropy is similar to that of the Alfv\'enic turbulence at large scales, and is consistent with the predictions of Section \ref{sec:kineticmodels} that kinetic turbulence remains anisotropic.

The two perpendicular electromagnetic waves that can exist in an isotropic $\beta\sim 1$ plasma for $1/\rho_\mathrm{i}\ll k_\perp\ll 1/\rho_\mathrm{e}$ are the kinetic Alfv\'en wave and the oblique whistler wave \citep[e.g.,][]{tenbarge12b,boldyrev13a} and the models in Section \ref{sec:kineticmodels} describe the nonlinear turbulence based on these modes. \citet{chen13c} developed a method to distinguish between these two types of turbulence, based on the relative level of density and magnetic fluctuations. The spectra in Figure \ref{fig:nbspectra} are of the normalised fluctuations,
\begin{equation}
\delta\tilde{n}=\sqrt{\frac{\beta_\mathrm{i}}{2}\left(1+\frac{T_\mathrm{e}}{T_\mathrm{i}}\right)\left[1+\frac{\beta_\mathrm{i}}{2}\left(1+\frac{T_\mathrm{e}}{T_\mathrm{i}}\right)\right]}\frac{\delta n}{n_0},\ \ \ \ \ \delta \tilde{\mathbf{b}}=\frac{\delta\mathbf{B}}{B_0}.
\label{eq:kawnorm}
\end{equation}
Similarly to the Alfv\'en ratio (Section \ref{sec:indices}), the kinetic Alfv\'en ratio can be defined as $r_\mathrm{KAW}=\delta\tilde{n}^2/\delta\tilde{b}_\perp^2$, which is $r_\mathrm{KAW}=1$ for a kinetic Alfv\'en wave (due to its pressure-balanced nature) and $r_\mathrm{KAW}\ll 1$ for an oblique whistler wave (due to its high frequency $\omega\gg k_\perp v_\mathrm{th,i}$). It can be seen from Figure \ref{fig:nbspectra} that in the range between ion and electron scales, the density and magnetic fluctuations are of similar amplitude, and $r_\mathrm{KAW}\sim 1$. This suggests that the turbulence is predominantly kinetic Alfv\'en in nature, with whistler fluctuations making up (at most) a small fraction, which is consistent with the transition from Alfv\'enic turbulence at larger scales. \citet{chen13c} measured the average kinetic Alfv\'en ratio to be $r_\mathrm{KAW}=0.75$ in the solar wind and $r_\mathrm{KAW}=0.79$ in a kinetic Alfv\'en turbulence simulation, indicating that while the turbulence follows the linear expectation to order unity, the nonlinearities introduce quantitative differences (see also Section \ref{sec:indices}).

The kinetic Alfv\'en nature of the turbulence is consistent with the flattening of the density spectrum seen to occur just before ion scales (e.g., Figure \ref{fig:nbspectra}). The flattening can be explained as the kinetic Alfv\'en component at small scales taking over from the compressive non-Alfv\'enic component at larger scales \citep{harmon05,chandran09c}, a model which is consistent with solar wind observations \citep{chen13a}. It is also consistent with a significant, rather than negligible, $\delta E_\|$ spectrum \citep{mozer13}, although the $\delta E_\|/\delta E_\perp$ ratio measured by \citet{mozer13} is larger than the linear prediction\footnote{This remains to be explained but may be either a nonlinear effect or related to the fact that the periods studied were unusually low amplitude, in order for $\delta E_\|$ to be obtained from a two-component electric field measurement.}. Finally, the measured ion scale spectral break can be compared to the expected dispersive scale for the transition from Alfv\'enic to kinetic Alfv\'en turbulence. \citet{chen14b} examined periods of extreme $\beta_\mathrm{i}$ so that the various ion scales could be distinguished, and found that for $\beta_\mathrm{i}\gg 1$ the break occurs close to the ion gyroscale ($k\rho_\mathrm{i}\sim 1$), which is indeed the dispersive scale for Alfv\'enic turbulence, but at $\beta_\mathrm{i}\ll 1$ it occurs at the ion inertial length ($kd_\mathrm{i}\sim 1$). This latter result remains to be explained, but one possibility is the presence of a significant $k_\|$ component in low-$\beta_\mathrm{i}$ turbulence \citep{boldyrev15}.

\subsection{Intermittency}

An important feature of turbulence, alluded to in Section \ref{sec:kineticmodels}, is intermittency, i.e., the non-Gaussian nature of the fluctuations. Intermittency has been well studied in the solar wind for $k\rho_\mathrm{i}<1$ \citep[see reviews by][]{bruno13,alexandrova13a}, where most quantities are seen to become increasingly non-Gaussian towards smaller scales, a well known feature of turbulence. This is related to the turbulent energy being concentrated into a smaller fraction of the volume as the cascade proceeds to smaller scales, and results in the formation of energetic structures within the plasma. Understanding this intermittency and structure generation is important for both the cascade process and plasma heating, since the energy dissipation is thought to be focussed at these structures.

\begin{figure}
\centering
\includegraphics[width=0.45\textwidth,trim=0 0 0 0,clip]{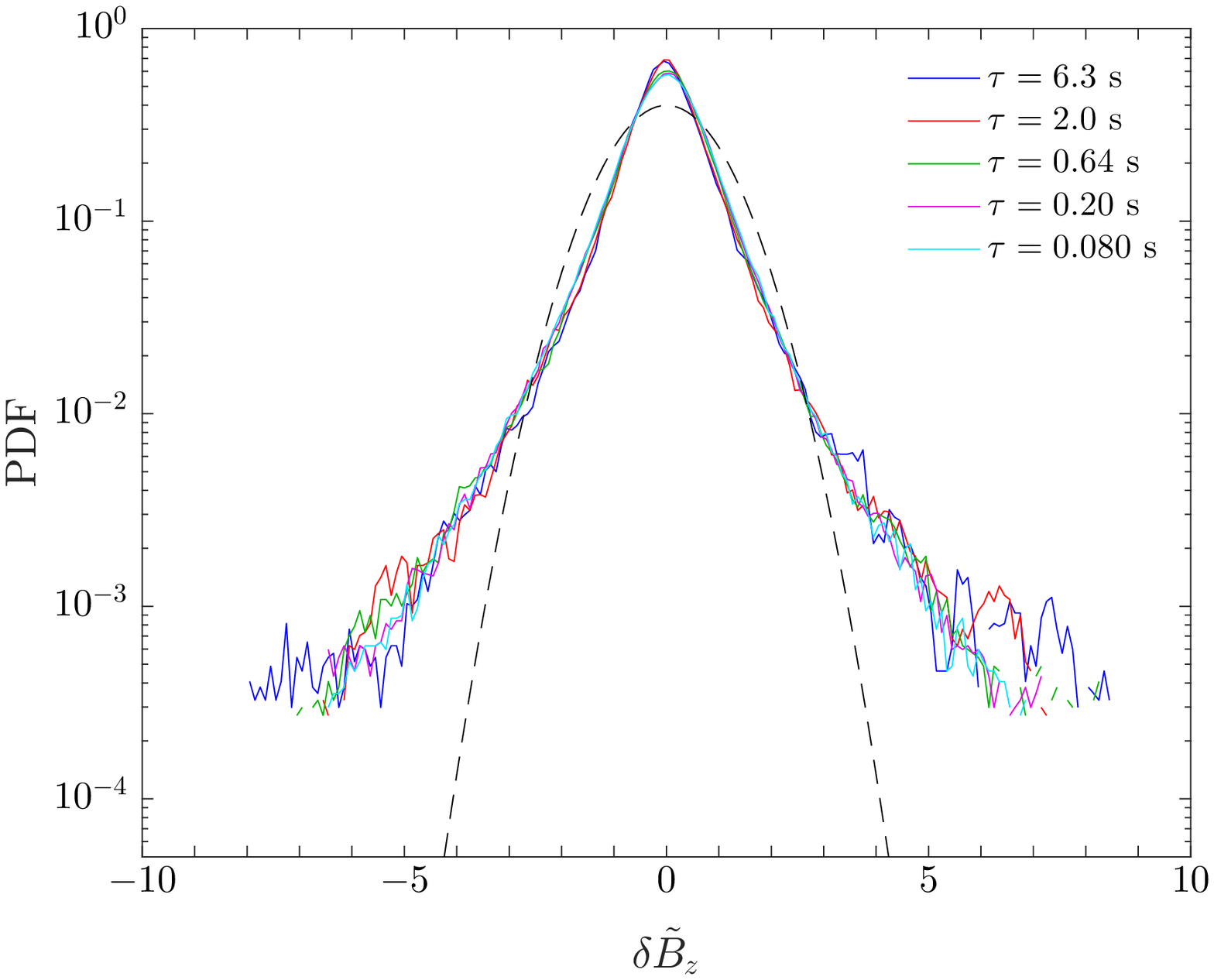}
\hspace{0.04\textwidth}
\includegraphics[width=0.45\textwidth,trim=0 0 0 0,clip]{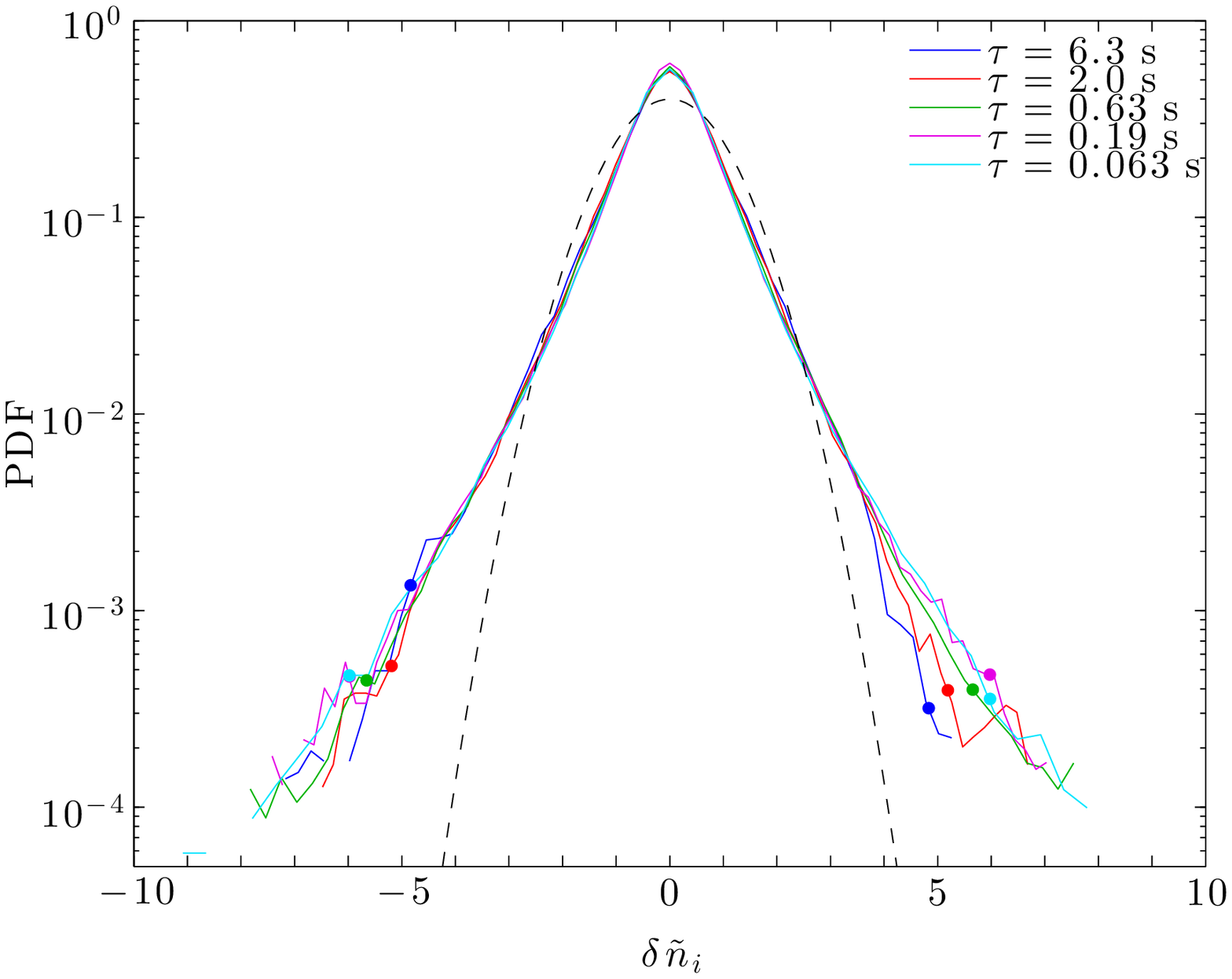}
\caption{Left: PDFs of magnetic fluctuations from ion to electron scales \citep[using the data interval of][]{chen10b}. Right: same for density fluctuations \citep[from][]{chen14a}. A Gaussian distribution is given by the black dashed line.}
\label{fig:intermittency}
\end{figure}

As a result of the Alfv\'enic cascade, the probability density functions (PDFs) of the fluctuations are significantly non-Gaussian by the time ion scales $k\rho_\mathrm{i}\sim 1$ are reached. Recently, it has been possible to investigate how the PDF shape continues to change down to electron scales. Intriguingly, \citet{kiyani09a} showed that from ion to electron scales, $1/\rho_\mathrm{i}<k<1/\rho_\mathrm{e}$, the rescaled PDFs of magnetic fluctuations do not get significantly more non-Gaussian, but retain a similar shape. \citet{chen14a} found a similar behaviour for the density fluctuation PDFs, which are shown in Figure \ref{fig:intermittency}, along with the magnetic fluctuations\footnote{Note that the density and magnetic fluctuations in this figure are from different time periods, which can account for the different absolute levels of non-Gaussianity.}. The similar behaviour of the density and magnetic fluctuations in the kinetic range is expected, since in kinetic Alfv\'en turbulence these fields are directly coupled (Section \ref{sec:kineticmodels}), although the self-similarity of the PDFs is unusual for a turbulent cascade. One possibility is that kinetic range turbulence is inherently mono-fractal; although a recent measurement of the multi-fractal spectrum by \citet{sorriso-valvo16} showed a larger multi-fractality in the the kinetic range. Another possibility is that the kurtosis of the fluctuations is limited by another process, such as the instability of the current sheet structures which may form \citep{biskamp90}. However, the nature of the intermittency in the kinetic range remains to be fully understood.

An alternative way to examine intermittency is through the distribution of rotation angles of the magnetic field. The rotation angle $\alpha$ is defined as the angle through which the field rotates over a given spatial scale (or time scale $\tau$ measured by a spacecraft under the Taylor hypothesis),
\begin{equation}
\alpha(t,\tau)=\cos^{-1}\left[\frac{\mathbf{B}(t)\cdot\mathbf{B}(t+\tau)}{|\mathbf{B}(t)||\mathbf{B}(t+\tau)|}\right],
\end{equation}
and has been used to study both the intermittency and the structures that are formed \citep[e.g.,][and references therein]{zhdankin12b}. Such analysis was recently extended into the kinetic range by \citet{chen15}, in which the distribution of $\alpha$ over scales $1/\rho_\mathrm{i}<k<1/\rho_\mathrm{e}$ was investigated. It was found that, similar to larger scales \citep{bruno04,zhdankin12b}, the PDFs of $\alpha$ are well-fit by log-normal distributions, with only small changes in shape from ion to electron scales.

\begin{figure}
\centering
\includegraphics[width=0.5\textwidth,trim=0 0 0 0,clip]{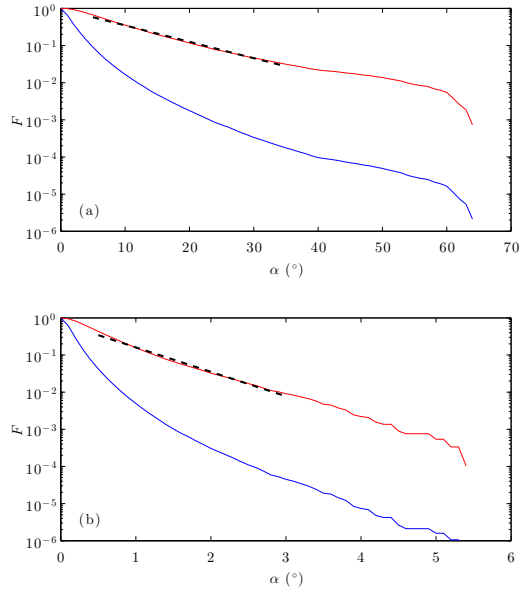}
\caption{Fraction of magnetic rotations larger than $\alpha$ (blue) and magnetic fluctuation energy in those angles (red) at (a) ion and (b) electron scales, along with exponential fits (black dashed) \citep[from][]{chen15}.}
\label{fig:rotations}
\end{figure}

The large rotations are sometimes identified as the ``structures'', so it is of interest to investigate their distribution, how much energy they contain, and how large the rotations are. Figure \ref{fig:rotations} shows the fraction of the time that the rotation angles are larger than $\alpha$, as a function of $\alpha$, for scales $\tau$ corresponding to the ion and electron gyroscales. Also shown is the fraction of magnetic fluctuation energy $|\delta\mathbf{B}|^2$ contained in these angles. It can be seen that for a given angle, the energy fraction is much larger than the filling fraction, consistent with the non-Gaussian nature of the fluctuations. Similar analyses have been performed for the energy dissipation at current structures in both MHD and kinetic simulations \citep{wan12a,wan16,zhdankin14,zhdankin16b}, with similar results. However, at kinetic scales, the absolute rotation angles are relatively small. The energy fraction in Figure \ref{fig:rotations} falls exponentially, with e-folding angle $9.8^\circ$ at ion scales and $0.66^\circ$ at electron scales, demonstrating that large angles $>30^\circ$ do not contain significant energy in the kinetic range. This places important constraints on energy conversion and dissipation mechanisms, such as magnetic reconnection, that have been proposed to be taking place here.

\section{Dissipation and future prospects}

While we have made significant progress in recent years, there are still many aspects of astrophysical plasma turbulence which remain to be understood. As well as the questions mentioned in the above sections, one of the major unsolved problems is how the turbulent energy is finally transferred to the particles and dissipated, i.e., how astrophysical plasmas, which are usually considered highly collisionless on the spatial and temporal scales of the turbulence, can be heated to the temperatures we observe. Work has begun on this question, but it will likely be from future measurements and new space missions that this problem is finally solved. In this Section, a selection of that recent work, and possible future directions, are discussed.

\subsection{Dissipation}
\label{sec:dissipation}

One basic question is how dissipation is distributed throughout a turbulent plasma, i.e., to what extent it is focussed at the energetic structures generated by intermittency. In fact, measuring the distribution of the dissipation can, in some sense, be considered a more direct way of measuring the intermittency of the cascade, since it is more directly connected to the energy cascade rate\footnote{Relating the fluctuations of the energy dissipation to those of the turbulent fields requires an additional assumption, i.e., the plasma turbulence equivalent of the \citet{kolmogorov62} refined similarity hypothesis \citep[see, e.g.,][]{chandran15}.}. \citet{zhdankin16a} measured the PDFs of the dissipation, averaged over different MHD-range scales, in both MHD simulations and the solar wind (where a proxy based on magnetic field measurements was used). 
The PDFs were found to be well-fitted by log-normal distributions in all cases\footnote{A log-normal was also found to fit the distribution of the local energy transfer rate in the solar wind \citep{sorriso-valvo15}.}, although the higher-order moments of both the solar wind and simulated distributions were better described by log-Poisson, rather than log-normal, scaling predictions. The rate at which these PDFs broaden towards smaller scales indicates the level of intermittency, and this was found to be consistent between the solar wind and simulations, and indicative of significant intermittency of the energy dissipation. It is of interest to determine which types of structures this dissipation is focussed at, and whether these correspond to any well-known plasma phenomena. For example, reconnecting current sheets \citep{retino07,sundkvist07}, Alfv\'en vortices \citep{alexandrova06}, double layers \citep{stawarz15}, and several others have been proposed. It remains to be determined which of these, if any, play a significant role in the dissipation of plasma turbulence.

A more fundamental, and even less well understood, question is the nature of the physical mechanisms which transfer energy from the electromagnetic turbulence to the particles and lead to the irreversible heating of the plasma. Several possibilities have been proposed, including ion cyclotron damping \citep[e.g.,][]{coleman68,smith12}, Landau damping \citep[e.g.,][]{howes08a,tenbarge13a}, stochastic heating \citep[e.g.,][]{chandran10b}, entropy cascade \citep{schekochihin09}, and reconnection associated mechanisms \citep[e.g.,][]{egedal12,drake14}. Identifying which combinations of these operate, under which conditions, will likely involve detailed examination of particle distributions. While future missions may be required to fully answer this (Section \ref{sec:futuremissions}), there have been some initial hints in the existing data. For example, \citet{he15a} interpreted the contours of measured ion distributions as being consistent with quasilinear expectations for the cyclotron and Landau resonances, and \cite{he15c} found plateaus in the distributions at the phase speeds of the Alfv\'en and slow mode waves, suggesting possible resonant damping of these modes. Various other indirect evidence has also been reported for some of the other mechanisms, although it will likely be from more direct techniques \citep[e.g.,][]{klein16b} that the answer to this question is finally determined.

\subsection{Future missions}
\label{sec:futuremissions}

In the coming years, we are set to continue understanding more about turbulence and heating in space and astrophysical plasmas, through several new missions which are due to be launched soon, or have been proposed. The \emph{Magnetospheric Multiscale} (\emph{MMS}) mission \citep{burch16}, launched in 2015, is already producing important new results, in particular from the increased time resolution of the ion and electron measurements \citep{pollock16short}, in the Earth's magnetosphere and magnetosheath. In 2017, the apogee of the \emph{MMS} orbit is due to be raised \citep{fuselier16}, so that for the extended mission, it will spend significant time in the free solar wind, allowing the study of kinetic turbulence there in even greater detail.

The following year (2018) is due to see the launch of \emph{Solar Probe Plus} \citep{fox15short} and \emph{Solar Orbiter} \citep{muller13}, which will both travel closer to the Sun than any previous spacecraft, and for which turbulence, along with the associated heating, is one of the main science goals. In particular, \emph{Solar Probe Plus} will travel to within 9 Solar radii of the solar surface, and will have a comprehensive payload measuring the electromagnetic fields \citep{bale16short} and particles \citep{kasper15short} to explore turbulence and heating within the corona for the first time. At the same time, the \emph{Voyager} spacecraft are moving further away from the Sun, allowing the first in situ measurements of turbulence in the local interstellar medium \citep{burlaga15}.

Finally, \emph{Turbulence Heating ObserverR} (\emph{THOR}) \citep{vaivads16short} is a mission concept proposed to the European Space Agency that is currently in study phase, and if selected would be due to launch in 2026. The mission will have the highest resolution set of instruments to date and will explore the range of near-Earth space environments to understand how turbulent energy is dissipated, how that energy is partitioned, and how this operates in different turbulence regimes. \emph{THOR} would be the first space mission dedicated to the topic of turbulent plasma heating, with broad implications for our understanding of astrophysical plasmas. Therefore, the coming years look very promising for the study of turbulence and heating in a variety of astrophysical environments.

\section*{Acknowledgements}

I would like to thank my collaborators who have worked with me on the projects discussed in this paper. I was supported by an STFC Ernest Rutherford Fellowship and an Imperial College Junior Research Fellowship.

\end{document}